\documentclass[a4paper]{cas-dc}

\usepackage{amsmath,amsfonts}
\usepackage{algorithmic}
\usepackage{algorithm}
\usepackage{array}
\usepackage[caption=false,font=normalsize,labelfont=sf,textfont=sf]{subfig}
\usepackage{textcomp}
\usepackage{stfloats}
\usepackage{booktabs}
\usepackage{multirow}
\usepackage{mathrsfs}
\usepackage{url}
\usepackage{graphicx}
\usepackage{hyperref}

\newcommand{\red}[1]{\textcolor{black}{#1}}

\usepackage[numbers]{natbib}
\def\tsc#1{\csdef{#1}{\textsc{\lowercase{#1}}\xspace}}
\tsc{WGM}
\tsc{QE}
\begin{document}
\let\WriteBookmarks\relax
\def\floatpagepagefraction{1}
\def\textpagefraction{.001}
\let\printorcid\relax % 可去掉页面下方的ORCID(s)

% Short title
% \shorttitle{<short title of the paper for running head>} 
\shorttitle{Two-and-a-half Order Score-based Model for Solving 3D Ill-posed Inverse Problems}    

% Short author
% \shortauthors{<short author list for running head>}
%\shortauthors{Zirong Li et al.}

% Main title of the paper
\title[mode = title]{Two-and-a-half Order Score-based Model for Solving 3D Ill-posed Inverse Problems}

\author[1]{Zirong Li}[]

\author[1]{ Yanyang Wang}[]

\author[1]{ Jianjia Zhang}[]

\author[1]{ Weiwen Wu}[]
\cormark[1] 
\author[2]{ Hengyong Yu}[]
\cormark[1]

\address[1]{Department of Biomedical Engineering, Sun-Yat-sen University, Shenzhen Campus, Shenzhen, China}
\address[2]{Department of Electrical and Computer Engineering, University of Massachusetts Lowell, Lowell, MA, USA}

\cortext[6]{Z. Li and Y. Wang contribute equally.} 
\cortext[1]{Corresponding authors are W. Wu (wuweiw7@mail.sysu.edu.cn) and H. Yu (hengyong-yu@ieee.org).} 

% Here goes the abstract
\begin{abstract}
Computed Tomography (CT) and Magnetic Resonance Imaging (MRI) are crucial technologies in the field of medical imaging. Score-based models \red{demonstrated effectiveness} in addressing different inverse problems encountered \red{in the field of} CT and MRI, such as sparse-view CT and fast MRI reconstruction. However, these models face challenges in achieving accurate three dimensional (3D) volumetric reconstruction. The existing score-based models \red{predominantly concentrate on reconstructing two-dimensional (2D) data distributions, resulting in inconsistencies} between adjacent slices in the reconstructed 3D volumetric images. To overcome this limitation, we propose a novel two-and-a-half order score-based model (TOSM). During the training phase, our TOSM learns data distributions in 2D space, \red{simplifying the training process compared to working directly on 3D volumes}. However, \red{during} the reconstruction phase, \red{the TOSM utilizes complementary scores along three directions (sagittal, coronal, and transaxial) to achieve a more precise reconstruction.} The development of TOSM is built on robust theoretical principles, ensuring its reliability and efficacy. Through extensive experimentation on large-scale sparse-view CT and fast MRI datasets, \red{our method achieved state-of-the-art (SOTA) results in solving 3D ill-posed inverse problems, averaging a 1.56 dB peak signal-to-noise ratio (PSNR) improvement over existing sparse-view CT reconstruction methods across 29 views and 0.87 dB PSNR improvement over existing fast MRI reconstruction methods with $\times$ 2 acceleration.} In summary, \red{TOSM significantly addresses the issue of inconsistency in 3D ill-posed problems by modeling the distribution of 3D data rather than 2D distribution,} which has achieved remarkable results in both CT and MRI reconstruction tasks.
\end{abstract}

% Use if graphical abstract is present
%\begin{graphicalabstract}
%\includegraphics{}
%\end{graphicalabstract}

% Research highlights
\begin{highlights}
\item We identify the issue of slice inconsistency in 3D volume reconstruction and develop an advanced score-based model to solve this problem.
\item We present an approach called Two-and-a-half Order Score-based Model (TOSM) for 3D reconstruction. Our method makes it feasible to apply a single 2D score-based model to 3D reconstruction.
\item We introduce groundbreaking theoretical derivations to validate the rationale and effectiveness of training a 2D score function for 3D volumetric images.
\item Our method has been successfully applied to reconstruct sparse-view CT and fast MRI, effectively addressing challenging ill-posed inverse problems.
\end{highlights}

% Keywords
% Each keyword is seperated by \sep
\maketitle

% Main text
\section{Introduction}
\label{sec:introduction}
Over the years, Computed Tomography (CT)\cite{guo2023spectral2spectral} and Magnetic Resonance Imaging (MRI)\cite{ranjbarzadeh2021brain} \red{have evolved as fundamental pillars in modern clinical diagnosis \red{for providing detailed information on organ, skeletal, and soft tissue structures.} \cite{chan2023attention}\cite{kumar2009light}\cite{aggarwal2002light}} Particularly, sparse-view CT \cite{li2023cascade} and fast MRI techniques \cite{chen2017low} rapidly achieve superior image quality \red{within reduced scanning durations  \cite{wu2021drone}.} Sparse-view CT imaging \red{typically decreases} scanning views and increases the view intervals to minimize \red{radiation exposure} and scanning time\cite{wu2022deep}. Similarly, fast MRI techniques aim to \red{diminish} measured data in k-space for faster scans\cite{yang2017dagan}. \red{Despite promising outcomes,} these advancements present new sets of ill-posed inverse problems \red{caused by incomplete measure data} demanding effective resolution\cite{chung2023solving}\cite{wu2021deep}.
\begin{figure}[ht]
    \centering
    \includegraphics[scale=0.4]{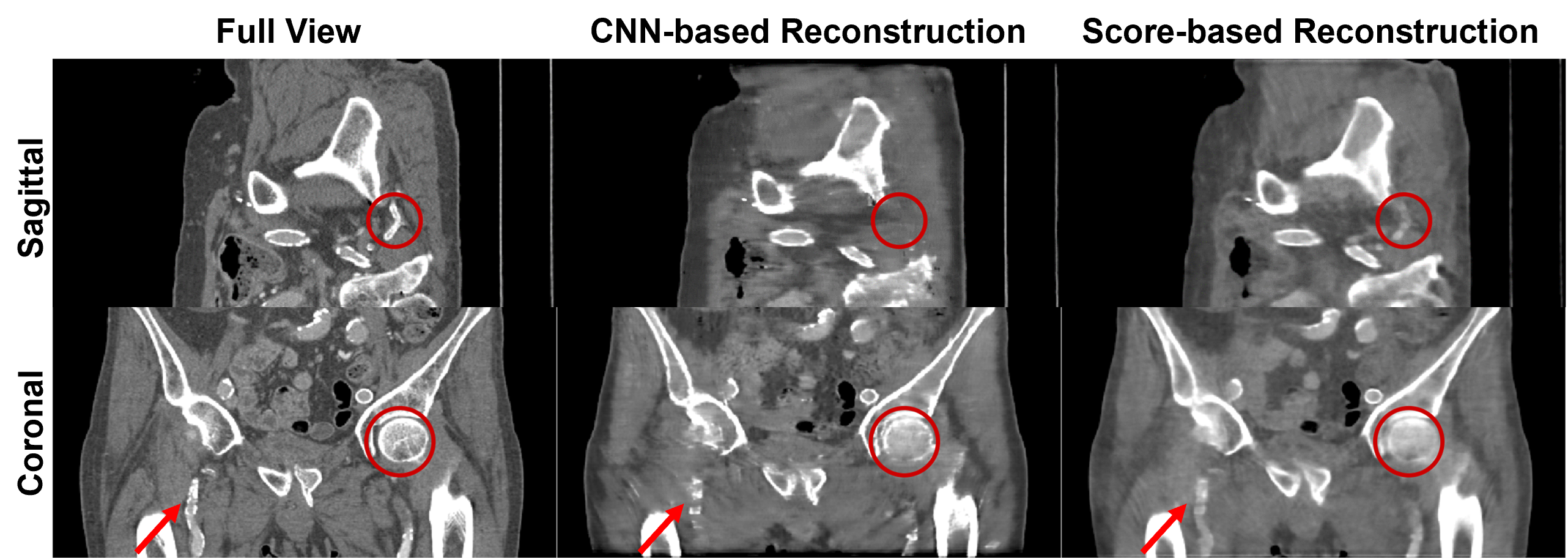}
    \caption{The reconstructed CT images from 29 views reveal noticeable inter-slice inconsistency in the sagittal and coronal planes when both the CNN-based and score-based reconstruction methods are employed.}
    \label{fig1}
\end{figure}

\red{In recent years, numerous advanced image processing and optimal methods have been proposed\cite{maini2018camera}\cite{chauhan2021experimental}\cite{ranjbarzadeh2022deep}}. Among them, \red{deep learning-based methods\cite{sun2022neural}\cite{haseli2023hecon}} have been developed to improve the reconstruction quality from individual slices within 2D space\cite{anari2022review}. \red{These methods include convolution neural networks (CNNs)\cite{gupta2018cnn}\cite{ranjbarzadeh2023me}\cite{tataei2021glioma}, generative adversarial networks (GANs)\cite{ying2019x2ct}, and more recent transformer techniques \cite{pan2022multi}\cite{kasgari2023point}.} Very recently, score-based models have emerged as a promising alternative approach to solving ill-posed inverse problems and achieved excellent results \cite{songsolving}\cite{songscore}. These models focus on learning the underlying data distribution and estimating the corresponding gradients. \red{By leveraging this gradient information, the score-based models can generate high-quality reconstructions from limited data or noisy measurements with measured data constraints\cite{chung2022score}\cite{guan2023generative}.}

However, \red{while most methods concentrate on reconstructing 2D image slices, their advancement falls short when aiming to achieve high-fidelity 3D volumetric images, as seen in sparse-view CT and fast MRI reconstruction \cite{xu2020sparse}.} When it comes to 3D volumetric reconstruction (see Fig. \ref{fig1}), the existing score-based models face a challenge known as inter-slice inconsistency\cite{chung2023solving}. This issue \red{arises from stacking independently reconstructed 2D slices to form a 3D volume\cite{reyneke2018review},} resulting in inconsistencies and artifacts, particularly in the sagittal and coronal planes. \red{This issue impedes the precise and coherent reconstruction of 3D volumetric images, a critical factor in clinical applications.\cite{mediouni2018review}\cite{mediouni2020translational}} 

The fundamental \red{challenge} of inter-slice inconsistency \red{arises from} the insufficient learning and modeling of the 3D data distribution \red{within} the score-based models \cite{khan2018methodological}. While 2D score-based models have demonstrated proficiency in capturing data distribution within 2D contexts, their limited capacity prevents them from extending this comprehension to cover the entire 3D volume. \red{Consequently}, the inter-slice relationships \red{crucial for achieving coherent and precise 3D reconstructions remain underutilized \cite{angelopoulou20153d}.} Indeed, it poses several challenges to training 3D score-based models for reconstruction tasks. Firstly, the \red{employing such models} in medical imaging demands substantial GPU resources to process 3D volumes. Secondly, it is a highly intricate process to train a 3D data distribution, often yielding suboptimal scores and consequently leading to \red{subpar} reconstructions. 

\red{In order to mitigate inter-slice inconsistency}, the state-of-the-art methods incorporate additional prior information to capture inter-slice relationships. For \red{instance}, one approach \red{involves implementing} the $L_1$ total variation (TV) regularization \cite{selesnick2012total} in the z-axis direction, \red{targeting the resolution} of discontinuities between slices\cite{chung2023solving}. Another method pre-trains two orthogonal score-based models to obtain inter-slice information\cite{lee2023improving}. \red{However, both of the methods mentioned earlier have limitations. Firstly, they lack complete modeling of the 3D data distribution or conduct score updates on 3D volumes.} Secondly, excessive regularization terms are introduced, leading to challenges in balancing multiple hyper-parameters. \red{Moreover, it incurs double computational costs for pre-training two models, limiting its practical deployment.}

To overcome the limitations in existing methods, \red{an optimal score-based model for 3D reconstruction should precisely model the distribution of 3D data}. Moreover, to release the computational burden associated with 3D scoring networks, \red{the reconstruction method should estimate the 3D distribution through a pre-trained 2D score-based model.} This can leverage the knowledge acquired from the transaxial plane, enabling higher computational efficiency for reconstruction while still benefiting from a comprehensive understanding of the 3D data distribution. Furthermore, to avoid introducing excessive hyper-parameters for fast convergence and lower complexity, a 2D score-based model can be \red{trained without extra regularization constraints to compute the 3D score}. This can simplify the training process and \red{improve the practicality and implementation of the reconstruction method in practice.}

Based on the aforementioned analysis, this paper presents a method called two-and-a-half (2.5) order score-based model (TOSM) to address the inter-slice inconsistency and leverage advanced score-based models for high-quality 3D reconstruction from under-sampled data. The critical concept of TOSM is based on the knowledge that the data distribution within each plane of the same volumetric image is similar. By training a 2D score-based model to learn the data distribution in the transaxial plane, we can utilize this model to calculate scores along the transaxial, sagittal, and coronal planes, resulting in three pseudo-3D scores for the reconstructed volumetric images. By providing evidence under specific conditions, we demonstrate that the actual 3D score of the volumetric image can be obtained by combining these three pseudo-3D scores, allowing for comprehensive gradient updates on the entire 3D image. Furthermore, to ensure the accuracy of the reconstructed 3D image, we incorporate 3D data consistency (e.g. SIRT-3D and K-space consistency) in the measurement space and effectively constrain the generation of our TOSM. During the reconstruction process, the powerful generative capability of the score-based model and the precise constraint of 3D data consistency complement each other, leading to remarkable results in solving the 3D ill-posed inverse problems. \red{This paper presents four significant contributions, which can be summarized as follows:}
\begin{itemize}
\item
We begin with an insightful observation. \red{Employing a sole pre-trained 2D score-based model enables the calculation of weighting scores across three orthogonal directions for an approximate estimation of the 3D score.} Remarkably, this approximation can be effectively determined through the combination of these three orthogonal directions.
\item
We present a groundbreaking approach to model the distribution of 3D volumetric images. By utilizing a single pre-trained 2D score-based model, we achieve a more accurate prediction of the 3D data distribution. \red{Additionally, to boost efficiency and accuracy, we introduce a computational strategy, leading to a substantial reduction in computational costs during the reconstruction process.} 
\item
We \red{provide} comprehensive theoretical derivations, that \red{validate} the rationale and effectiveness of training a 2D score function for 3D volumetric images. For the CT reconstruction task, we incorporate the Simultaneous Iterative Reconstruction Technique (SIRT) \red{as part of the data consistency criteria to enhance the reconstruction process.} The exceptional convergence of TOSM ensures stable and deterministic reconstruction results.
\item
Our method is successfully \red{employed in the reconstruction of medical volumetric images, effectively tackling challenging ill-posed inverse problems like sparse-view CT and fast MRI reconstruction.} The results are truly remarkable, \red{which means that from a clinical perspective, the radiation exposure to patients can be reduced, and the scanning time can be further shortened, alleviating the discomfort caused by prolonged scanning.}
\end{itemize}

\red{The subsequent sections of this paper are structured as follows. In Section II, we present an overview of medical 3D ill-posed inverse problems with existing solving methods and the fundamentals of score-based models. Section III elaborates on our motivation and gives the theory of our proposed TOSM method with detailed implementation. In Section IV, we present the experimental studies and evaluate the performance of our method on 3D sparse-view CT and fast MRI reconstruction tasks. Lastly, we delve into related discussions and conclude this paper in the final section.}
\section{Related Works}
\red{In recent years, deep learning methods have been widely applied in the field of medical imaging\cite{al2024fundus}\cite{mohammed2023hybrid} and some of them have already achieved excellent results\cite{mukhlif2023classification}\cite{mukhlif2023incorporating}. In this section, we will first introduce medical ill-posed inverse problems and some of the existing methods for solution. Subsequently, we will discuss the mathematical theory of novel denoising score matching models (score-based models) and their application in solving 3D inverse problems. Finally, we will discuss the research gaps in existing methods.}
\subsection{Medical Ill-posed Inverse Problems}
The fundamental challenge in \red{the field of} medical image reconstruction lies in \red{the resolution of} ill-posed inverse problems. The imaging problem can be \red{mathematically expressed} as $\boldsymbol{\mathcal{P}} = \mathcal{M}(v)$, where $v$ \red{corresponds to} the \red{object being imaged}, $\mathcal{M}$ models the measurement process, and $\mathcal{P}$ is the measured data generated from $v$. In the context of CT imaging, $\boldsymbol{\mathcal{P}}$ represents the sinogram, while in MRI, $\mathcal{P}$ represents the k-space data. The inverse problem involves finding $\mathcal{M}^{-1}$ such that $v=\mathcal{M}^{-1}(\mathcal{P})$. \red{However, this inverse problem is \red{commonly} ill-posed \red{as a result of} under-sampling during the data acquisition process.} Therefore, it is crucial to develop effective methods to address such ill-posed inverse problems.
\subsubsection{Sparse-view CT Reconstruction}
Classical sparse-view CT reconstruction methods \red{can be divided into two distinct categories}. The first category consists of the traditional methods that employ regularization techniques to promote the sparsity of the reconstructed images. For example, compressed sensing (CS) algorithms leverage the assumption that natural images are sparse in specific domains such as wavelet or total variation \cite{liu2015median}\cite{liu2015reconstruction}. These methods assume that the underlying image can be well-represented using only a few coefficients in a suitable transform domain. \red{The second category includes deep learning-based methods, which have emerged as a powerful alternative.} Approaches based on Convolutional Neural Networks (CNNs) (e.g., FBPConvNet \cite{jin2017deep}, Transformer-based MIST \cite{pan2022image}, and score-based methods \cite{song2020improved}) have shown promising results in sparse-view CT reconstruction.
\subsubsection{Fast MRI Reconstruction}
\red{Fast MRI reconstruction has become a topic of great interest in recent years, with researchers proposing various approaches to accelerate the imaging process and improve the quality of reconstructed images}\cite{chance1998novel}. One commonly explored direction is to leverage sparsity in MR images. CS-based approaches have been successfully applied to accelerate MRI by acquiring under-sampled k-space data and utilizing prior knowledge that MR images are sparse in certain domains \cite{lustig2008compressed}. These methods reconstruct high-quality images from significantly reduced data measurements. Additionally, deep learning-based methods have shown promising results. \red{CNNs and Generative Adversarial Networks (GANs) \cite{shende2019brief} have been trained to learn the mapping between under-sampled k-space data and fully-sampled images,} enabling fast and accurate reconstruction. More recently, the score-based models have also demonstrated excellent results.
\subsection{Denoising Score Matching Models}
Score-based models are \red{a type of} probabilistic generative models that employ a forward diffusion process and a backward denoising process during training \cite{song2020improved} \cite{wavelet}.
\red{During the forward process, the noise of different scales is incrementally added to the input image,} progressively degrading the image until it becomes pure Gaussian noise. In the backward generation process, the noise in the image is iteratively removed, effectively sampling the image from random Gaussian white noise. At each iteration, a neural network (e.g. U-Net) is employed to estimate the current noise. In the context of medical imaging, \red{the inverse problem involves the reconstruction on the probability density distribution $p(\textbf{x})$ of the reconstructed image $\textbf{x}$ \cite{songsolving}.}
\subsubsection{Noise Conditioned Score Networks (NCSNs)}
\red{The score function of the data probability density $p(\textbf{x})$ is formally defined as the logarithmic gradient of the data probability density}, $\nabla_{\textbf{x}} \log p(\textbf{x})$. The Langevin annealing algorithm is utilized to move the initial random probability density $p(\textbf{x}_0)$ towards regions of high probability density \cite{song2019generative}. The gradient of the logarithmic density serves as a force to guide the probability density into high-density regions. The iterative update equation is as follows:
\begin{equation}
\setlength\abovedisplayskip{0.1cm}
\setlength\belowdisplayskip{0.1cm}
\textbf{x}_{i}=\textbf{x}_{i-1}+\frac{\rho}{2} \nabla_{\textbf{x}_{i-1}} \log p(\textbf{x}_{i-1})+\sqrt{\rho} \cdot \omega_{i} ,
\end{equation}
where  $i \in{(1, \ldots, N)}$, $\rho$ controls the magnitude of the update in the direction of the score, $\textbf{x}_0$ is sampled from a prior distribution, and $\omega_{i} \sim \mathcal{N}(0, \mathbf{I})$ represents random disturbance. Consequently, the scoring model $s_{\theta}(\textbf{x}) \approx \nabla_{\textbf{x}} \log p(\textbf{x})$ can be trained to predict the derivatives of the probability density gradient, and the training loss function is defined as:
\begin{equation}
\setlength\abovedisplayskip{0.1cm}
\setlength\belowdisplayskip{0.1cm}
    \mathcal{L}_{\mathrm{sm}}=\mathbb{E}_{\textbf{x} \sim p(\textbf{x})}\left\|s_{\theta}(\textbf{x})-\nabla_{\textbf{x}} \log p(\textbf{x})\right\|_{2}^{2}.
    \label{eqloss}
\end{equation}
\subsubsection{Stochastic Differential Equations (SDEs)}
\red{In the context of  Stochastic Differential Equation Score Models (SDEs),} the diffusion and restoration processes are treated as continuous operations, which are solved by using stochastic differential equations. The forward diffusion process of SDEs \red{can be} defined as \red{follows}:
\begin{equation}
\setlength\abovedisplayskip{0.1cm}
\setlength\belowdisplayskip{0.1cm}
    \partial \textbf{x}=\left[\mathcal{F}(\textbf{x}, t)-\sigma(t)^{2} \cdot \nabla_{\textbf{x}} \log p_{t}(\textbf{x})\right] \cdot \partial t+\sigma(t) \cdot \partial \hat{\omega},
\end{equation}
where $\mathcal{F}$ is a function of $\textbf{x}$ and $t$ to compute the drift coefficients, $\sigma$ is a time-dependent function that computes diffusion coefficients, $\hat{\omega}$ represents the Brownian motion, and $\sigma(t)^{2} \cdot \nabla_{\textbf{x}} \log p_{t}(\textbf{x})$ acts as the force to drive the probability density for high-density regions. Similarly to NCSNs, a neural network $s_{\theta}(\textbf{x}, t) \approx \nabla_{\textbf{x}} \log p_{t}(\textbf{x})$ can be trained to estimate scores, and the loss function is defined as:
\begin{equation}
\setlength\abovedisplayskip{0.1cm}
\setlength\belowdisplayskip{0.1cm}
    \mathcal{L}_{\mathrm{sm}}=\mathbb{E}_{\textbf{x} \sim p(\textbf{x})}\left\|s_{\theta}\left(\textbf{x}_{t}, t\right)-\nabla_{\textbf{x}_{t}} \log p_{t}\left(\textbf{x}_{t} \mid \textbf{x}_{0}\right)\right\|_{2}^{2}.
\end{equation}
In practice, small steps are used instead of continuous quantities, and an iterative solution is as follows:
\begin{equation}
\setlength\abovedisplayskip{0.1cm}
\setlength\belowdisplayskip{0.1cm}
\textbf{x}_{t-1} = \textbf{x}_{t} + \left[\mathcal{F}(\textbf{x}, t) - \sigma(t)^{2} \cdot \nabla_{\textbf{x}} \log p_{t}(\textbf{x})\right] \cdot \Delta t + \sigma(t) \cdot \Delta \hat{\omega}.
\end{equation}
\subsubsection{3D Score-based Model}
\red{The utilization of} a 3D score-based model \red{involves expanding} the scoring network from a 2D U-Net to a 3D U-Net architecture, \red{allowing for} score computation \red{within 3D space}. However, \red{the computational challenge escalates} significantly owing to \red{the substantial volume of voxels, particularly} when the data dimension exceeds $64^3$ \cite{liu2019point}. Besides, Chung \emph{et. al.} mentions that it is difficult to apply \red{3D score-based models} to address medical imaging issues\cite{chung2023solving}. As a result, the current applications of 3D score-based models primarily \red{focus on}  tasks \red{like} point cloud denoising, small voxel generation, and structural prediction \cite{luo2021diffusion}, rather than high-resolution voxel-based 3D reconstruction.
\red{\subsubsection{Research Gaps}
Through the above introduction of score models and the discussion in the introduction section about their shortcomings in solving the problem of 3D reconstruction, we can observe a research gap in precisely mathematically modeling and reconstructing high-resolution 3D volumes using score models. Existing methods either model only 2D data distributions or cannot handle high-resolution large-scale medical 3D volumes.}
\section{Methodology}
\subsection{TOSM Motivation}
\begin{figure}[t]
    \centering
    \includegraphics[scale=0.37]{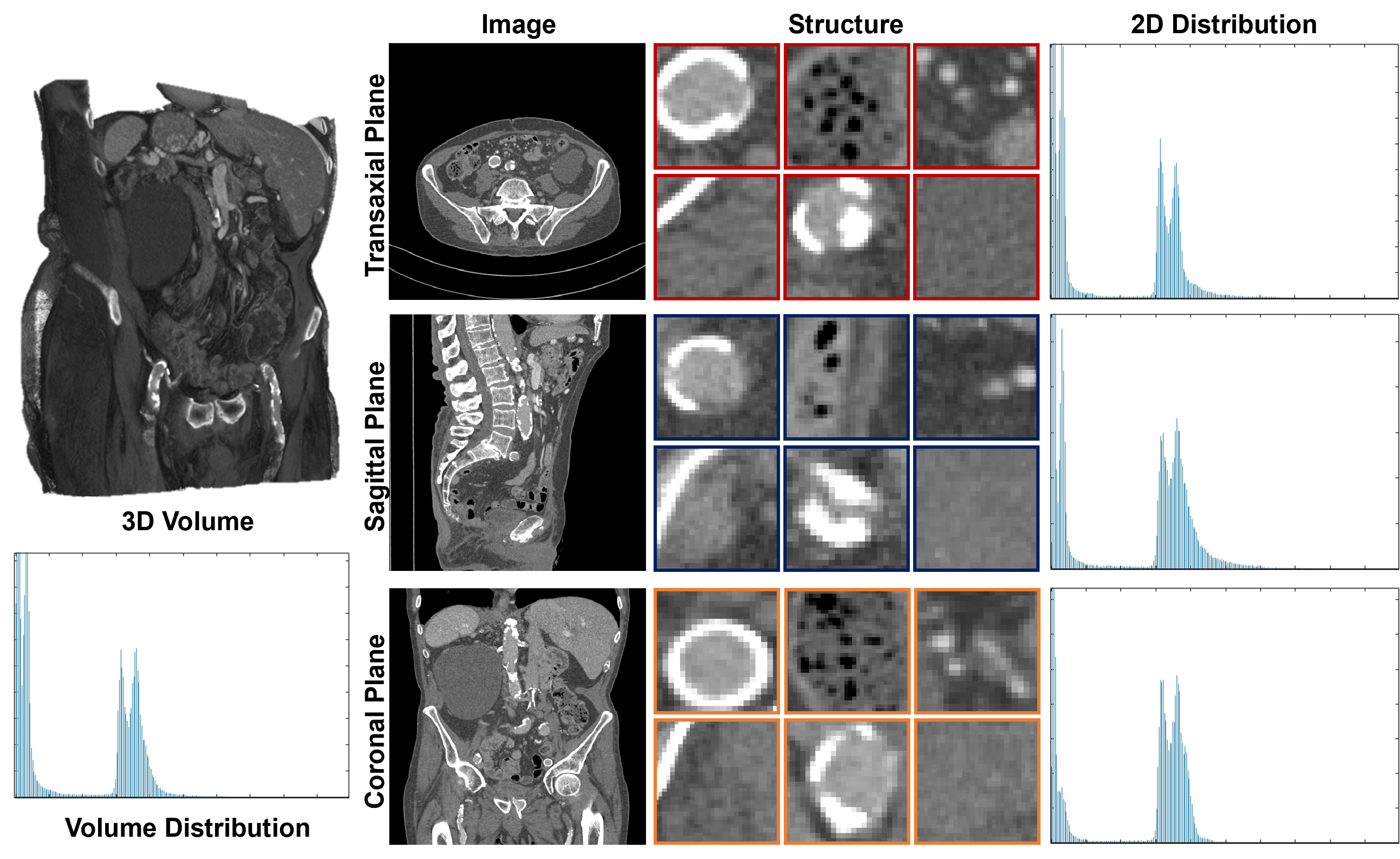}
    \caption{Similarity among the histogram distribution of 3D volumetric images and the corresponding 2D distributions of slice images along three directions.}
    \label{intro2}
\end{figure}

For a 3D volumetric image from one patient, \red{intriguing observations emerge}. As shown in Fig. \ref{intro2}, the left panel \red{displays} a 3D volumetric image \red{along with its} corresponding histogram distribution. The middle panel \red{exhibits} transaxial, sagittal, and coronal views of the 3D image with enlarged regions of interest (ROIs). The ROIs in different views exhibit strikingly similar fine structural details, \red{demonstrating} a high level of \red{consistency} among slices from \red{diverse} planes (see the right panel). This shared structural coherence across \red{distinct views} underscores the consistency and integrity of the 3D volumetric data. \red{Furthermore, upon scrutinizing histogram distributions of image slices from the three different planes, significant similarities become apparent.} Moreover, the histogram distributions of slices \red{within} the three planes closely resemble the histogram distribution of the entire 3D volumetric image. This consistency in histogram distributions \red{solidifies the concept of shared characteristics} and overall coherence among slices in the volumetric image.

\red{The observations in the preceding context lead us to propose a hypothesis.} Due to the striking similarity in feature structures and distributions of slices across different directions, the distributions of slices within the 3D volume can be approximated as equivalent from the point of view of big data. Consequently, \red{employing a single neural network trained on 2D slices to model the 3D data distribution becomes a viable approach, efficiently harnessing shared information across diverse planes.} These findings \red{mark a significant stride} in 3D reconstruction, as they highlight the inherent coherence and consistency of the data in different plane views. \red{Leveraging these structural and content similarities can streamline the reconstruction process, resulting in more efficient and robust outcomes. This advancement holds promise for} enhancing the progress in medical imaging and other fields that rely on 3D volumetric reconstruction techniques.
\subsection{TOSM Overview}
\begin{figure*}[t]
    \centering
    \includegraphics[scale=0.53]{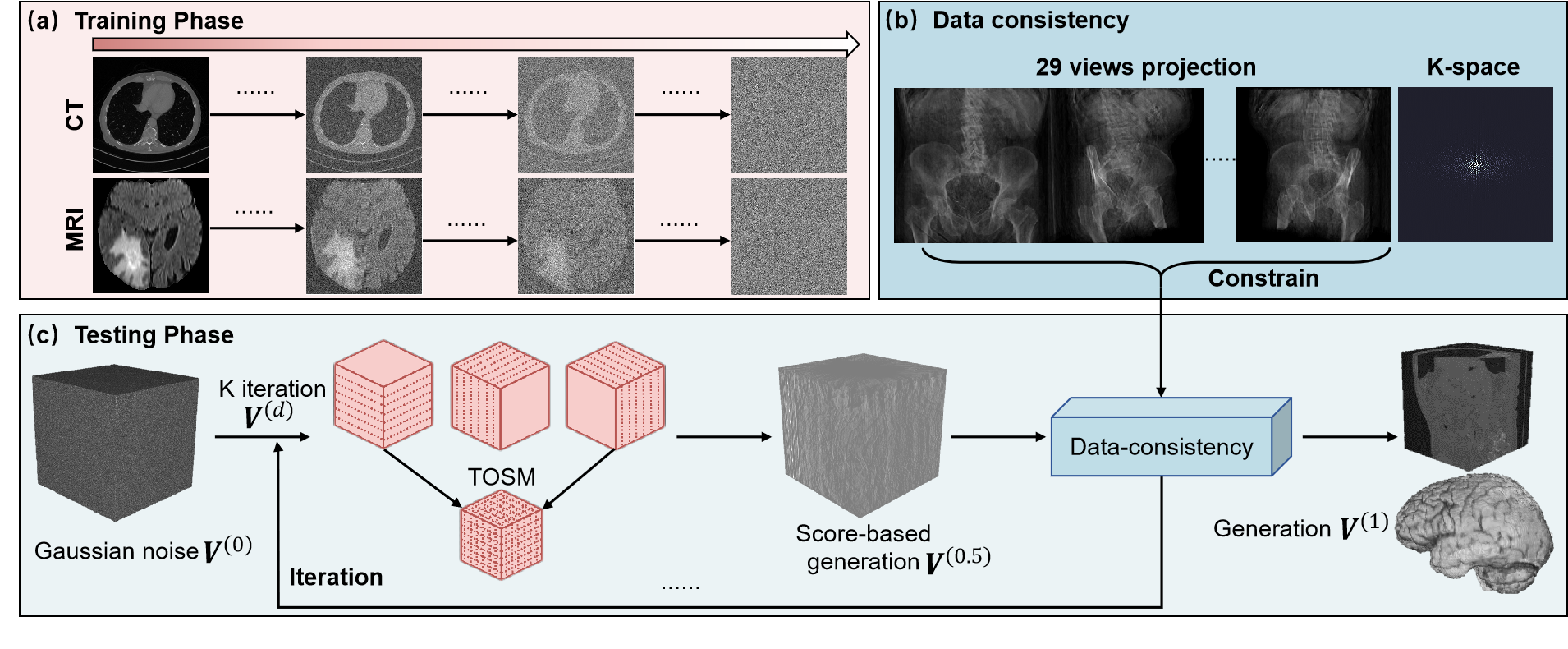}
    \caption{ The TOSM method is employed to solve the inverse problem through a well-defined pipeline. Initially, 3D Gaussian noise is iteratively generated using a 2.5-order score model. Subsequently, a data consistency constraint is introduced. By iteratively combining these two processes, high-quality 3D reconstruction is achieved.}
    \label{pipline}
\end{figure*}

\begin{figure}[t]
    \centering
    \includegraphics[scale=0.23]{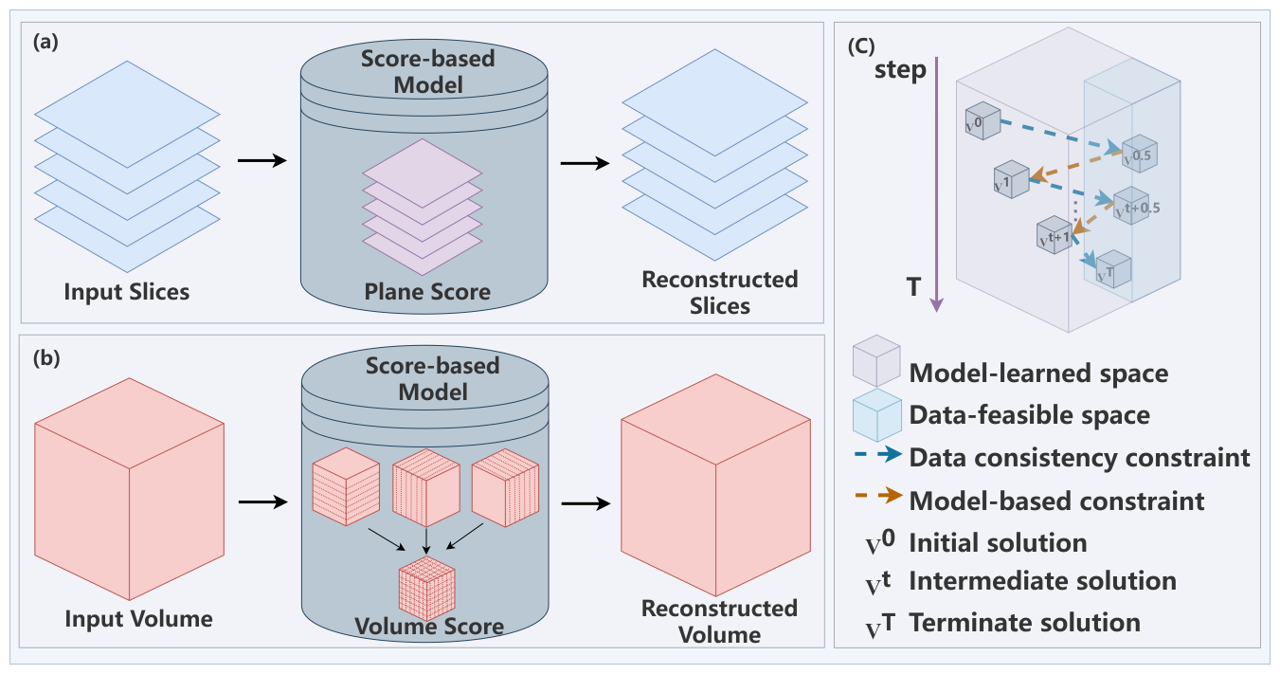}
    \caption{ Comparison of the traditional 2D score-based models and our TOSM for 3D Reconstruction. (a) Existing 2D score-based model for 3D reconstruction, where the 3D volumetric image is sliced into 2D slices and subsequently stacked together. (b) The process of 3D volumetric reconstruction using TOSM: the entire volume is inputted into the model. The pre-trained 2D score-based model calculates scores in three mutually orthogonal directions. (c) The optimization process of TOSM.}
    \label{compare}
\end{figure}

\red{The successful application of score-based models for 3D reconstruction presents a significant challenge to accurately recover the data distribution of 3D volumetric images while accounting for noise.} Inspired by the observed similarity property of 3D volumetric images, we present a pioneering solution of the 2.5-order score-based model (TOSM). As shown in Fig. \ref{pipline}, our TOSM method builds upon conventional score-based models by training the scoring function to learn 2D distributions from 2D slices during the training phase.
However, it is essential to highlight that relying solely on scores obtained from the coronal plane slice-by-slice, as typically done in the 2D score-based model during the reconstruction process (see Fig. \ref{compare}), is an approximation. To overcome this limitation, our TOSM introduces a novel approach for computing 3D scores. During the reconstruction process, our TOSM leverages the trained scoring network to score each layer of the 3D volumetric image along the sagittal, coronal, and transaxial planes. \red{This yields three 3D scores that are precisely aligned with the dimensions of the reconstructed volumetric image.}
\red{However, a challenge arises as these 3D scores alone cannot fully represent the actual scores independently due to the lack of inter-slice information.} To address this issue, TOSM adopts a weighted sum of these 3D scores, meticulously considering the spatial relationships and dependencies between adjacent slices. This approach enables us to obtain an authentic 3D score for the reconstructed volumetric image, as the weighting strategy incorporates 3D scores while taking into account the inter-slice connections from three different planes. In essence, the proposed TOSM approach \red{effectively addresses} the limitations of conventional score-based models, significantly enhancing the accuracy of 3D reconstruction by effectively incorporating vital inter-slice information and comprehensively considering the spatial context of the volumetric images.
\subsection{TOSM Theory}
The primary challenge in utilizing the score-based models for 3D reconstruction \red{lies in accurately modeling the data distribution of 3D volumetric images from noise.} In this paper, we present an approach called 2.5-order score-based model (TOSM) to address this challenge. To begin, we consider three mutually orthogonal directions in 3D space: $x\in \mathbb{R}$, $y\in \mathbb{R}$, and $z\in \mathbb{R}$. We define the 3D volumetric image as $\mathbf{v}(x,y,z)\in \mathbb{R}$, where $\mathbf{v}(x,y,z)$ represents an element of the 3D volumetric image at $(x,y,z)$. \red{In the following, we omit the position dependency and denote the element as $ \mathbf{v}$ when there is no ambiguity.} Our first task is to model the probability density, $p(\mathbf{v})$, of the 3D volumetric images. For this purpose, we define $g_x(y,z)\in \mathbb{R}^2  $, $h_y(x,z)\in \mathbb{R} ^2$, and $f_x(x,y)\in \mathbb{R} ^2$ as the 2D distributions of the $y-z$, $x-z$, and $x-y$ planes, respectively. The Dirac function $\delta(\cdot)$ plays a crucial role in the modeling process, and we define the data distribution of the volumetric images from three different directions as follows:
\begin{equation}
\setlength\abovedisplayskip{0.1cm}
\setlength\belowdisplayskip{0.1cm}
    p(\mathbf{v}) = \int _{-\infty} ^{+\infty }\delta (x-w)g_{x=w}(y,z)dw
    \label{eq1},
\end{equation}
\begin{equation}
\setlength\abovedisplayskip{0.1cm}
\setlength\belowdisplayskip{0.1cm}
    p(\mathbf{v}) = \int _{-\infty} ^{+\infty }\delta (y-w)h_{y=w}(x,z)dw
    \label{eq2},
\end{equation}
\begin{equation}
\setlength\abovedisplayskip{0.1cm}
\setlength\belowdisplayskip{0.1cm}
    p(\mathbf{v}) = \int _{-\infty} ^{+\infty }\delta (z-w)f_{z=w}(x,y)dw
    \label{eq3}.
\end{equation}
Combining these three approaches, we express the distribution $p(\mathbf{v})$ as follows:
\begin{equation} 
\setlength\abovedisplayskip{0.1cm}
\setlength\belowdisplayskip{0.1cm}
    \begin{aligned}
  \left [  p(\mathbf{v}) \right ]^{\alpha+\beta+\gamma}= \left [ \int _{-\infty} ^{+\infty }\delta (x-w)g_{x=w}(y,z)dw \right ] ^\alpha 
\\ \times \left [ \int _{-\infty} ^{+\infty }\delta (y-w)h_{y=w}(x,z)dw \right ] ^\beta \\ \times \left [ \int _{-\infty} ^{+\infty }\delta (z-w)f_{z=w}(x,y)dw \right ] ^\gamma.
    \end{aligned}
    \label{eq4}
\end{equation}
\red{Due to the observed similarity in the distributions of 2D slices across different planes within a single volume, it is possible to approximate these distributions using a common function. Consequently, we propose the following approximation:}
\begin{equation} 
\setlength\abovedisplayskip{0.1cm}
\setlength\belowdisplayskip{0.1cm}
    \begin{aligned}
  \left [  p(\mathbf{v}) \right ]^{\alpha+\beta+\gamma}= \left [ \int _{-\infty} ^{+\infty }\delta (x-w)f_{x=w}(y,z)dw \right ] ^\alpha 
\\ \times \left [ \int _{-\infty} ^{+\infty }\delta (y-w)f_{y=w}(x,z)dw \right ] ^\beta \\ \times \left [ \int _{-\infty} ^{+\infty }\delta (z-w)f_{z=w}(x,y)dw \right ] ^\gamma.
    \end{aligned}
    \label{eq5}
\end{equation}

To obtain scores, which correspond to the gradients of the logarithmic data distribution, we simultaneously take the logarithm and gradient on both sides of (\ref{eq5}) and \red{establish the hyper-parameters in such a way that} $\alpha + \beta + \gamma = 1$:
\begin{equation} 
\setlength\abovedisplayskip{0.1cm}
\setlength\belowdisplayskip{0.1cm}
    \begin{aligned}
\bigtriangledown _\mathbf{v} \log_{}{}p(\mathbf{v}) =\alpha  \bigtriangledown _{y,z} \log_{}{}\left [ \int _{-\infty} ^{+\infty }\delta (x-w)f_{x=w}(y,z)dw \right ]
\\ + \beta  \bigtriangledown _{x,z} \log_{}{}\left [ \int _{-\infty} ^{+\infty }\delta (y-w)f_{y=w}(x,z)dw \right ]
\\ + \gamma  \bigtriangledown _{x,y} \log_{}{}\left [ \int _{-\infty} ^{+\infty }\delta (z-w)f_{z=w}(x,y)dw \right ].
    \end{aligned}
    \label{eq6}
\end{equation}
In practical applications, $x$, $y$, and $z$ are discrete variables. Using the Kronecker delta function $\delta^{'}$, we can discretize (\ref{eq6}) as follow:
\begin{equation} 
\setlength\abovedisplayskip{0.1cm}
\setlength\belowdisplayskip{0.1cm}
    \begin{aligned}
\bigtriangledown _\mathbf{v} \log_{}{}p(\mathbf{v}) =\alpha  \bigtriangledown _{y,z} \log_{}{}\sum_{k=0 }^{K} \left [ \delta^{'}_{xk}f^{}_{x=k}(y,z)\right ]
\\ + \beta  \bigtriangledown _{x,z} \log_{}{}\sum_{j=0 }^{J} \left [ \delta^{'}_{yj} f^{}_{y=j}(x,z)\right ]
\\ + \gamma  \bigtriangledown _{x,y} \log_{}{}\sum_{q=0}^{Q} \left [ \delta^{'}_{zq} f^{}_{z=q}(x,y)\right ],
    \end{aligned}
    \label{eq7}
\end{equation}
 where $K$, $J$, and $Q$ \red{denote the slice numbers of the $y-z$, $x-z$, and $x-y$ planes, respectively.} Since the 2D data distributions of each plane are subsets of the same 3D distribution in our proposed TOSM, 3D scores can be predicted using a single pre-trained 2D network, denoted as $S_\theta$, where the network parameters $\theta$ remain the same. Consequently, we can reach the following equation (\ref{eq8}), \red{where $\mathbf{v}_t$ represents an initial 3D random noise obeying a Gaussian normal distribution, and $t$ denotes the total number of diffusion steps:}
\begin{equation} 
\setlength\abovedisplayskip{0.1cm}
\setlength\belowdisplayskip{0.1cm}
    \begin{aligned}
\bigtriangledown _{\mathbf{v}_{t}} \log_{}{}p(\mathbf{v}_t) &\approx \alpha \sum_{k=0}^{K}S_{\theta}(\mathbf{v}_{t}[k,:,:],t)
+ \beta \sum_{j=0}^{J}S_{\theta}(\mathbf{v}_{t}[:,j,:],t)
\\ &+ \gamma \sum_{q=0}^{Q}S_{\theta}(\mathbf{v}_{t}[:,:,q],t).
    \end{aligned}
    \label{eq8}
\end{equation}
By employing the above transformations, the 3D score problem is converted into the computation of 2D scores in three directions. \red{Consequently, our proposed TOSM can utilize a trained 2D score-based model to compute the overall 3D scores.}

\subsection{TOSM Applications}
As shown in Fig. \ref{pipline}(a), \red{In the training phase, through utilizing high-quality transaxial plane data as the training set, the score-based model can train a neural network, denoted as $S_{\theta}$ with equation (\ref{eqloss}). This network enables the efficient computation of the scoring function, essentially allowing estimation of the reciprocal of the probability density function. Consequently, it achieves modeling of the 2D data distribution.}
In the testing phase, as illustrated in Fig. \ref{pipline}(c), TOSM can effectively address ill-posed inverse problems in medical imaging. The key to solve an ill-posed inverse problem is to find the solution of $\mathbf{v}=\mathcal{M}^{-1}(\boldsymbol{\mathcal{P}})$, where $\mathcal{M}$ is an operator to generate the measured data $\boldsymbol{\mathcal{P}}$, and $\mathbf{v}$ is the scanned object. Due to the ill-posed nature of this problem, we adopt a regularized iterative approach:
\red{
\begin{equation}
\setlength\abovedisplayskip{0.1cm}
\setlength\belowdisplayskip{0.1cm}
    \boldsymbol{\mathbf{v}}^{*}=\underset{\mathbf{v}}{\operatorname{argmin}} \frac{1}{2}\|\mathcal{P}-\mathcal{M} \mathbf{v}\|_{2}^{2}+\frac{\eta}{2}R(\mathbf{v}),
\end{equation}
}
where $R{(\cdot)}$ is a suitable regularization, and $\eta$ is the regularization factor to balance the data fidelity and regularization prior. Here, we use a score from the 2.5-order score-based model as the regularization term to guide the optimizing process,
\red{
\begin{equation}
\setlength\abovedisplayskip{0.1cm}
\setlength\belowdisplayskip{0.1cm}
    \mathbf{v}^{*}=\underset{\mathbf{v}}{\operatorname{argmin}} \frac{1}{2}\|\boldsymbol{\mathcal{P}}-\mathcal{M} \mathbf{v}\|_{2}^{2}+\frac{\eta}{2}\bigtriangledown _\mathbf{v} \log_{}{}p(\mathbf{v}).
    \label{eq15}
\end{equation}
}
As illustrated in Fig. \ref{compare}, the volumetric image generated by the score-based model resides in the space learned by the model, and \red{it consists of the feasible data space.  The solution} ${\mathbf{v}}^{d+1}$ can be iteratively updated in two steps:
\red{
\begin{equation}
\setlength\abovedisplayskip{0.1cm}
\setlength\belowdisplayskip{0.1cm}
    \mathbf{v}^{d+0.5}=\underset{\mathbf{v}}{\operatorname{argmin}} \frac{1}{2}\|\mathbf{v}- \boldsymbol{\mathbf{v}^{d}}\|_{2}^{2}+\frac{\eta}{2}\bigtriangledown _\mathbf{v} \log_{}{}p(\mathbf{v}),
    \label{eq16}
\end{equation}
}
%%%%
\red{
\begin{equation}
\setlength\abovedisplayskip{0.1cm}
\setlength\belowdisplayskip{0.1cm}
     \mathbf{v}^{d+1}=\underset{\mathbf{v}}{\operatorname{argmin}} \frac{1}{2}\|\boldsymbol{\mathcal{P}}-\mathcal{M} \mathbf{v}\|_{2}^{2}+\frac{\psi}{2}\|\mathbf{v}- \boldsymbol{\mathbf{v}^{d+0.5}}\|_{2}^{2},
    \label{eq17}
\end{equation}
}
where the symbol $\psi$ serves as a factor and $d$ represents the current iteration number. \red{Equation (\ref{eq16}) represents a score model-based regularization prior that can be updated using the trained 3D scores.} To tackle (\ref{eq17}) in sparse-view CT and fast MRI reconstruction tasks, we employ different iteration strategies as data consistency measures to constrain the reconstruction process. Specifically, for sparse-view CT reconstruction, we utilize the classical SIRT method to solve (\ref{eq17}). Conversely, in MRI reconstruction tasks, we adopt the k-space consistency to enforce constraints on the reconstruction process.  Through iterative updates of the reconstructed volumetric image, we enhance the consistency between the generated volumetric image and the observed data measurements. \red{Through the incorporation of these data consistency measures, the TOSM framework can effectively resolve the ill-posed inverse problems encountered in medical imaging, thereby contributing to improved reconstruction accuracy and robustness.}
%\begin{equation}
%    \boldsymbol{v}^{d+1}=\lambda SIRT(\boldsymbol{v}^{d})+\mu [\boldsymbol{v}^{d}-\bigtriangledown _{v^d} \log_{}{}p(\boldsymbol{v}^{d})],
%    \label{eq15}
%\end{equation}
%By iteratively updating the reconstructed volume based on the acquired sparse-view data, we improve the consistency between the generated volume and the observed data.

%\begin{equation}
%    \boldsymbol{v}^{d+1}=\lambda consistency_{k-space}(\boldsymbol{v}^{d})+\mu [\boldsymbol{v}^{d}-\bigtriangledown _{v^d} \log_{}{}p(v^d)],
%    \label{eq15}
%\end{equation}
%where $\lambda,\mu$ are constants and detailed information about $consistency_{k-space}$ will be provided later in the text. By considering the acquired k-space data and the corresponding inverse Fourier transform, we enforce consistency between the generated volume and the measured k-space data, leading to accurate and reliable MRI reconstruction results. 
\subsubsection{Sparse-view CT Reconstruction}
In the sparse-view CT reconstruction task, the available measured views in the projection domain are sparse and incomplete, resulting in sparse-view artifacts. Inspired by the iterative reconstruction method SIRT\cite{gregor2008computational}, we want to generate a current volumetric image that aligns with the original projections from each view. \red{To achieve this objective, we define initial parameters for ensuring data consistency.} Specifically, $p_a$ represents a projection value for $a^{th}$ x-ray path, $\mathbf{v}_n$ denotes the $n^{th}$ pixel of $\mathbf{v}$, and $N$ represents the number of pixels within $\mathbf{v}$. $m_{an}$ represents a weight factor, indicating the contribution of $\mathbf{v}_n$ for $a^{th}$ ray. We \red{initiate the process} by calculating the estimated projection value for the x-ray path $a$:

\begin{equation}
\begin{array}{l}
p_{a}^{*}=\sum_{n=1}^{N} m_{a n} \mathbf{v}_{n}^{(d)},
\end{array}
\end{equation}
%Then calculate the error between the actual projection and the estimated projection:
%\begin{equation}
%\begin{array}{l}
%\Delta_{a}=y_{a}-y_{a}^{*}
%\end{array}
%\end{equation}
where $b$ represents the set of x-ray paths across all projection views. Subsequently, we modify the value of the $b$ part using the following equation:
\begin{equation}
\begin{array}{l}
C_{b}=\sum_{a \in b} \frac{p_{a}-\sum_{n=1}^{N} m_{a n} v_{n}^{(d)}}{\sum_{n=1}^{N} m_{a n}} m_{a b}.
\end{array}
\end{equation}
Finally, the SIRT is used to update by the following expression:
\begin{equation}
\begin{array}{l}
\mathbf{v}_{n}^{(d+1)}=\mathbf{v}_{n}^{(d)}+ \frac {C_{b}}{\textstyle \sum_{a \in b} m_{a n}}^{},
\end{array}
\end{equation}
and it can be further written as
\begin{equation}
\begin{array}{l}
\mathbf{v}_{n}^{(d+1)}=\mathbf{v}_{n}^{(d)}+ \frac{\sum_{a \in b} \frac{p_{a}-\sum_{n=1}^{N} m_{a n} \mathbf{v}_{n}^{(d)}}{\sum_{n=1}^{N} m_{a n}} m_{a n}}{\sum_{a \in b} m_{a n}} .
\end{array}
\end{equation}

\subsubsection{Fast MRI Reconstruction}
In fast MRI reconstruction, we simplify the calculation by dividing the data consistency into two parts: one in the image domain and the other in K-space. The data consistency is defined as follows:
\begin{equation}
{\mathbf{v}}^{(d+1)}=\left({I}-\gamma_1 {M_2}^{*} {~M_2}\right) {\mathbf{v}^{(d)}}+{M_2}^{*} \mathcal{P^{*}},
\label{eq18}
\end{equation}
where $\gamma_1 \in [0,1]$, ${M_2}^{*}$ is the Hermitian conjugate of $M_2$, ${M_2}$ and $ {M_2}^{*}$ represent the 2D discrete Fourier forward and inverse transformations, and $\mathcal{P^{*}}$ represents the measured data in k-space.
\red{Following the framework proposed by Chung \emph{et. al.}\cite{chung2022score},} the data consistency constraints need to satisfy a non-extended mapping property. The details on the pertaining to Eq. (\ref{eq18}) can refer to \cite{tang2011projection}.

\red{In the context of MRI reconstruction, it is imperative to consider the complex nature of the data as well as the phase information.} Unlike CT reconstruction, MRI data cannot be directly processed using neural networks\cite{hyun2018deep}. To address this, we divide the data into real and imaginary parts, treating them in different channels. This approach effectively \red{enhances} the stability of the reconstruction process\cite{chung2022score}.
\section{Experimental Setup and Results}
\begin{figure*}[t]
    \centering
    \includegraphics[scale=0.7]{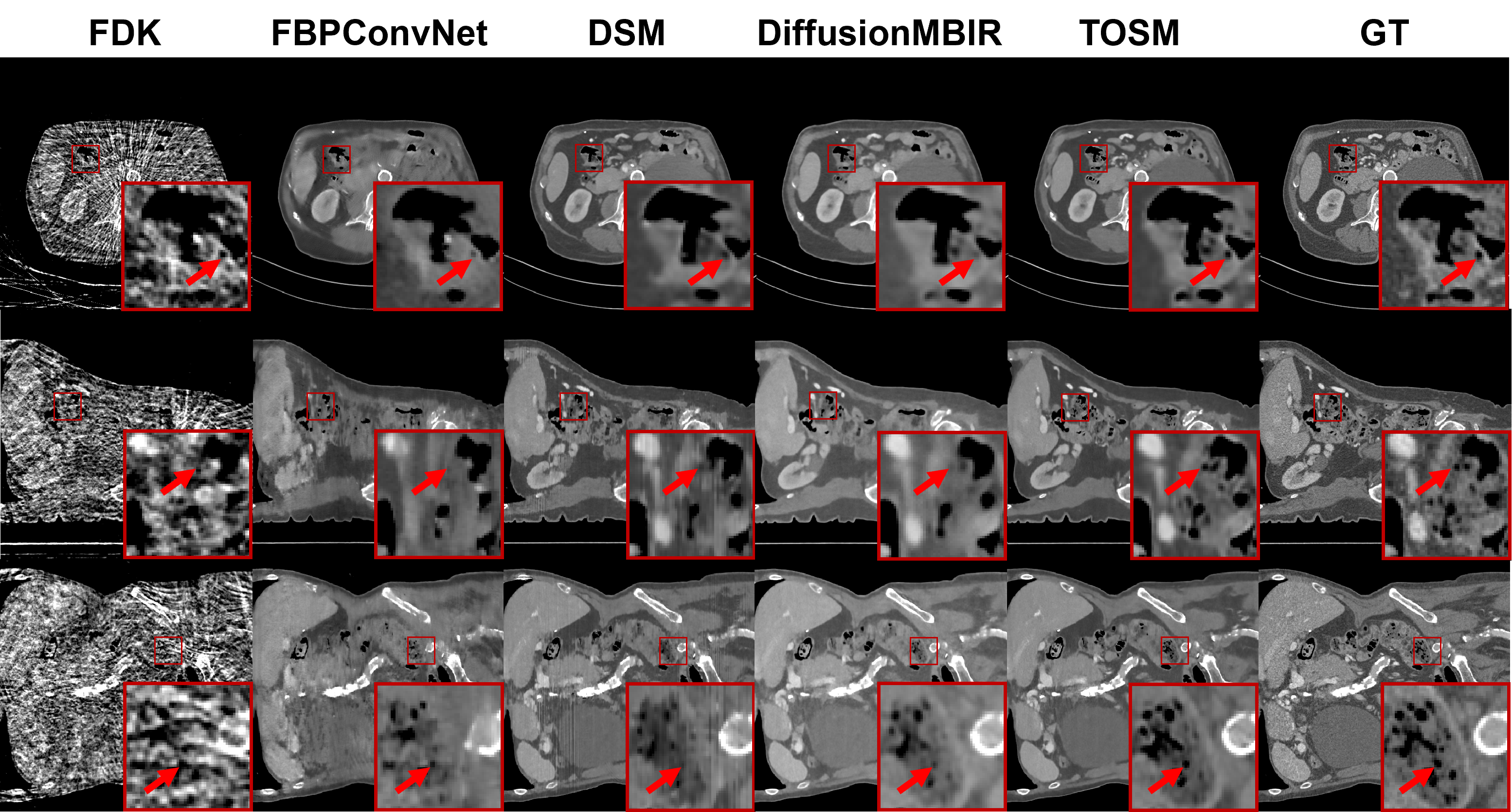}
    \caption{Representative reconstructed results of cone-beam CT from 29 views by different methods.  From top to bottom, the rows correspond to the transaxial, sagittal, and coronal planes. The first to fifth columns correspond to different reconstruction methods, and the last column shows the ground truth (GT) images.}
    \label{resultCT}
\end{figure*}
%%%%%%%%%%%%%%%%%%
%%%%%%%%%%%%%%%%%%
\begin{figure}[ht]
    \centering
    \includegraphics[scale=0.32]{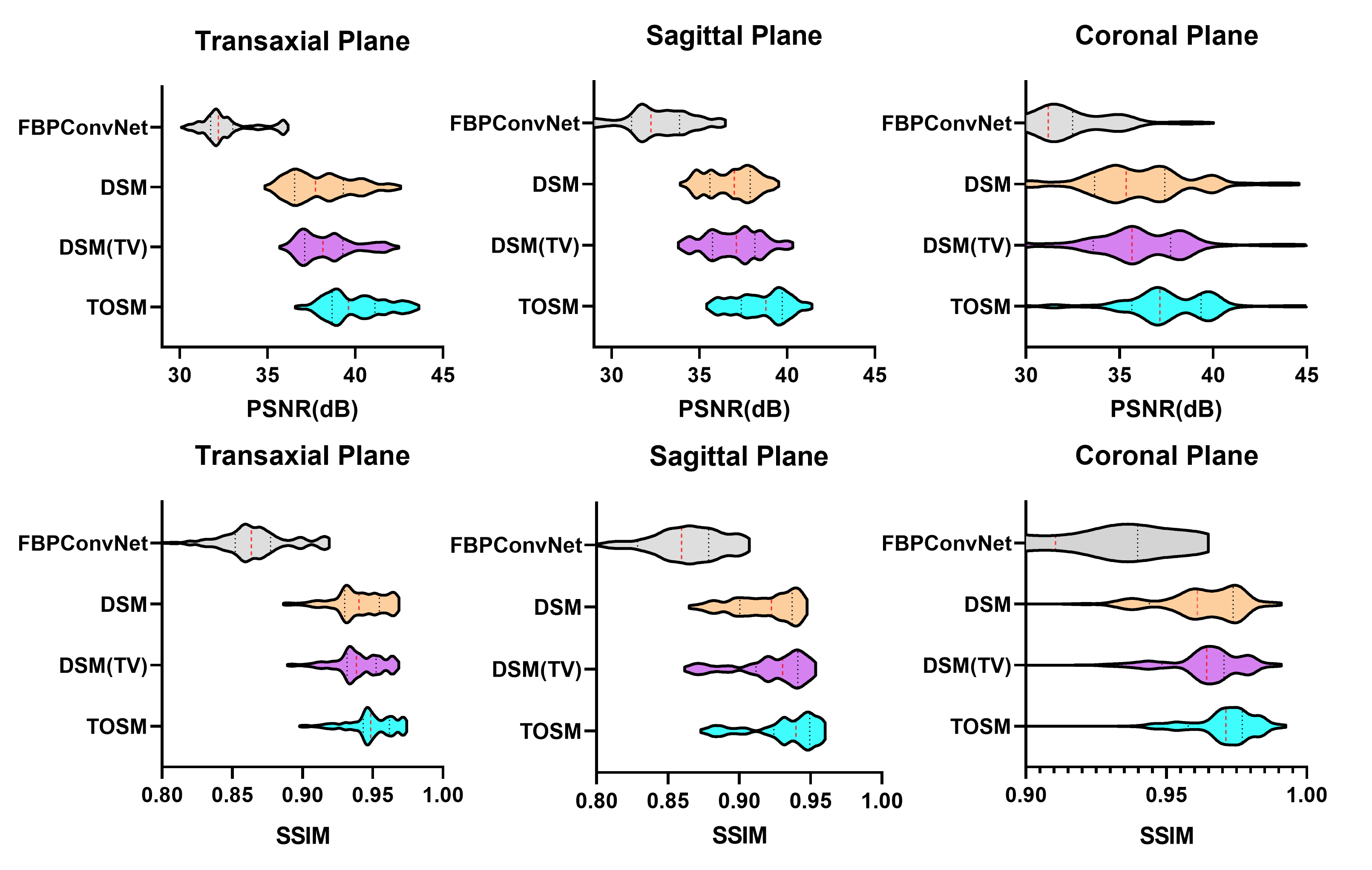}
    \caption{The violin plot for the statistical results of different methods, including FBPConvNet, DSM, DiffusionMBIR, and our TOSM. It is obvious that our TOSM achieves the best results.}
    \label{violin}
\end{figure}

\begin{figure}[ht]
    \centering
    \includegraphics[scale=0.28]{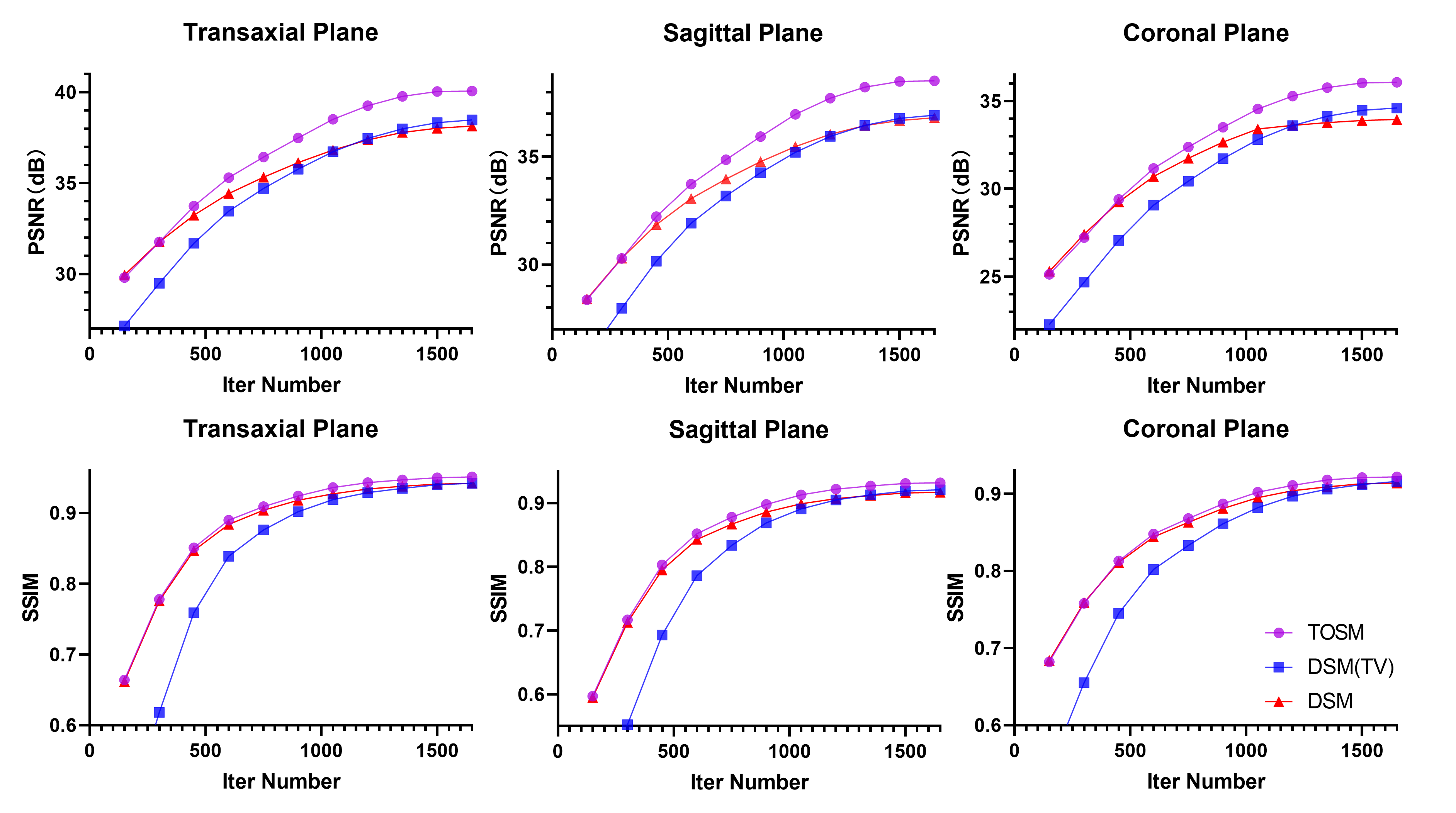}
    \caption{The convergence plots of PSNR and SSIM with respect to the number of iterations in the testing phase.}
    \label{converge}
\end{figure}
\begin{figure*}[t]
    \centering
    \includegraphics[scale=0.7]{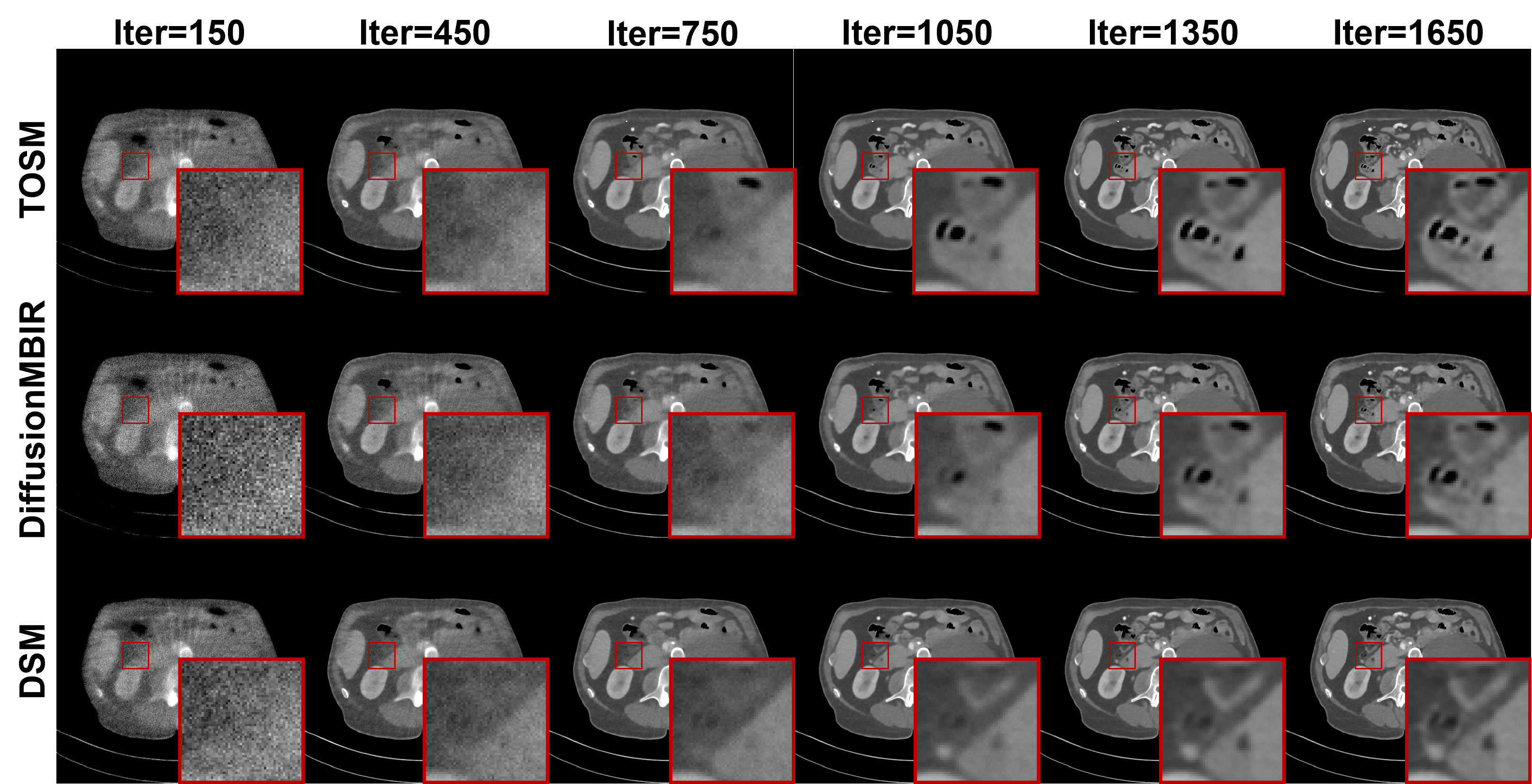}
    \caption{Reconstructed images of three scoring models with respect to different iteration numbers. Clearly, TOSM outperforms the other scoring model methods at any iteration number.}
    \label{image_iter}
\end{figure*}
\begin{figure}[t]
    \centering
    \includegraphics[scale=0.42]{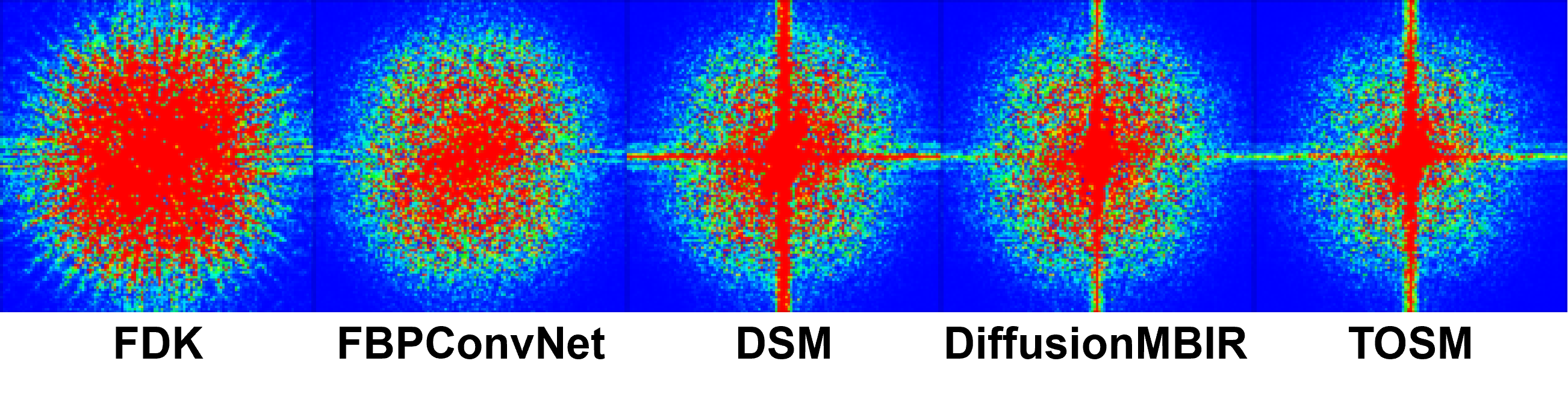}
    \caption{The noise power spectra (NPS) plots of different methods. The central region of the plots represents low-frequency noise components, while the outer region represents high-frequency noise components.}
    \label{NPS}
\end{figure}
\begin{figure}[ht]
    \centering
    \includegraphics[scale=0.24]{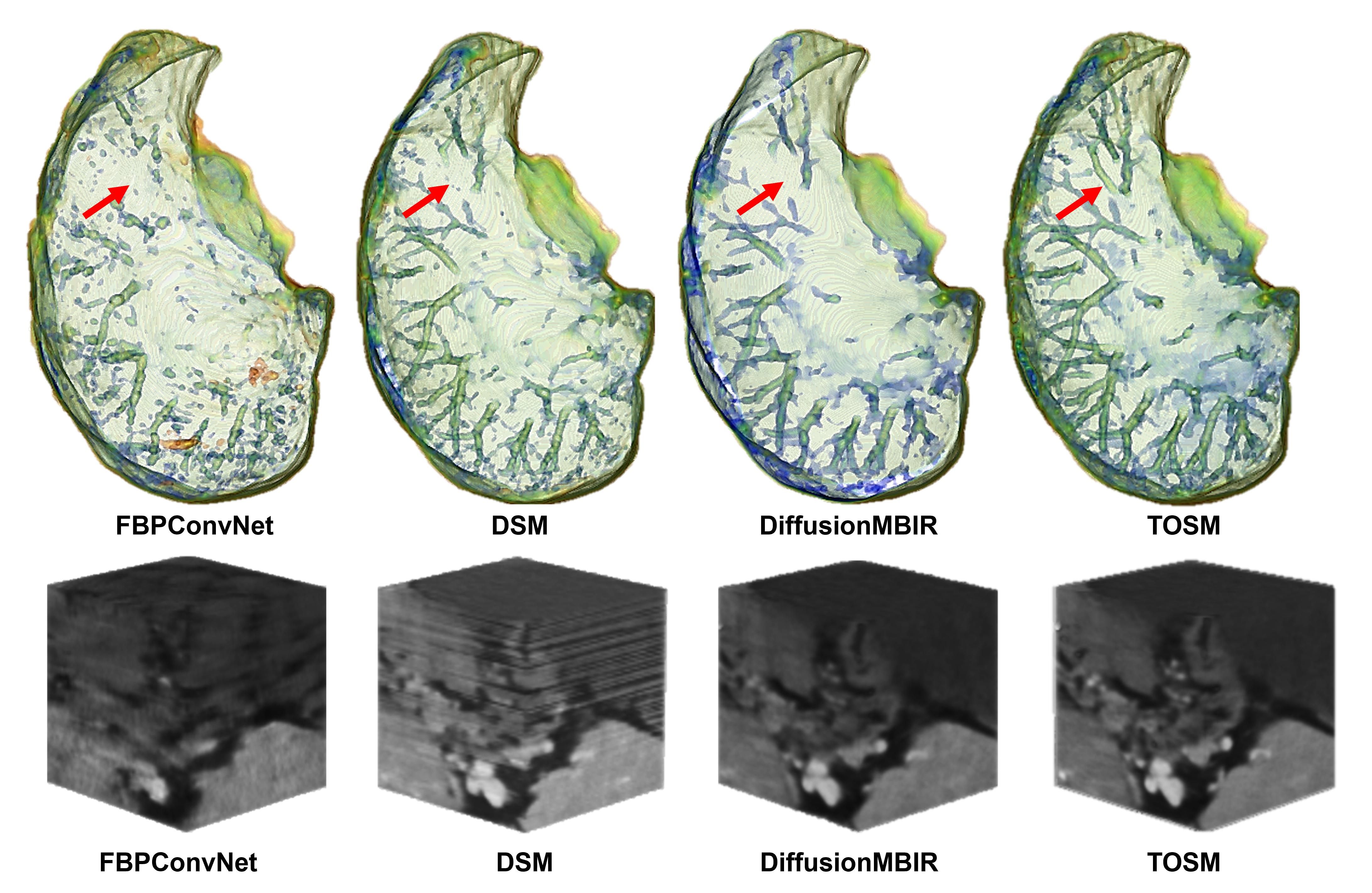}
    \caption{Visualization of representative 3D volumetric reconstruction results. The top row is 3D volumetric renderings of the lung CT from different methods, where the tubular structures represent pulmonary blood vessels and the red arrows indicate notable features. The bottom row are 3D ROI rendering of 100$\times$100$\times$100 volumetric images.}
    \label{fig3d}
\end{figure}
%%%%%%%%%%%%%%%
\begin{figure*}[t]
    \centering
    \includegraphics[scale=0.7]{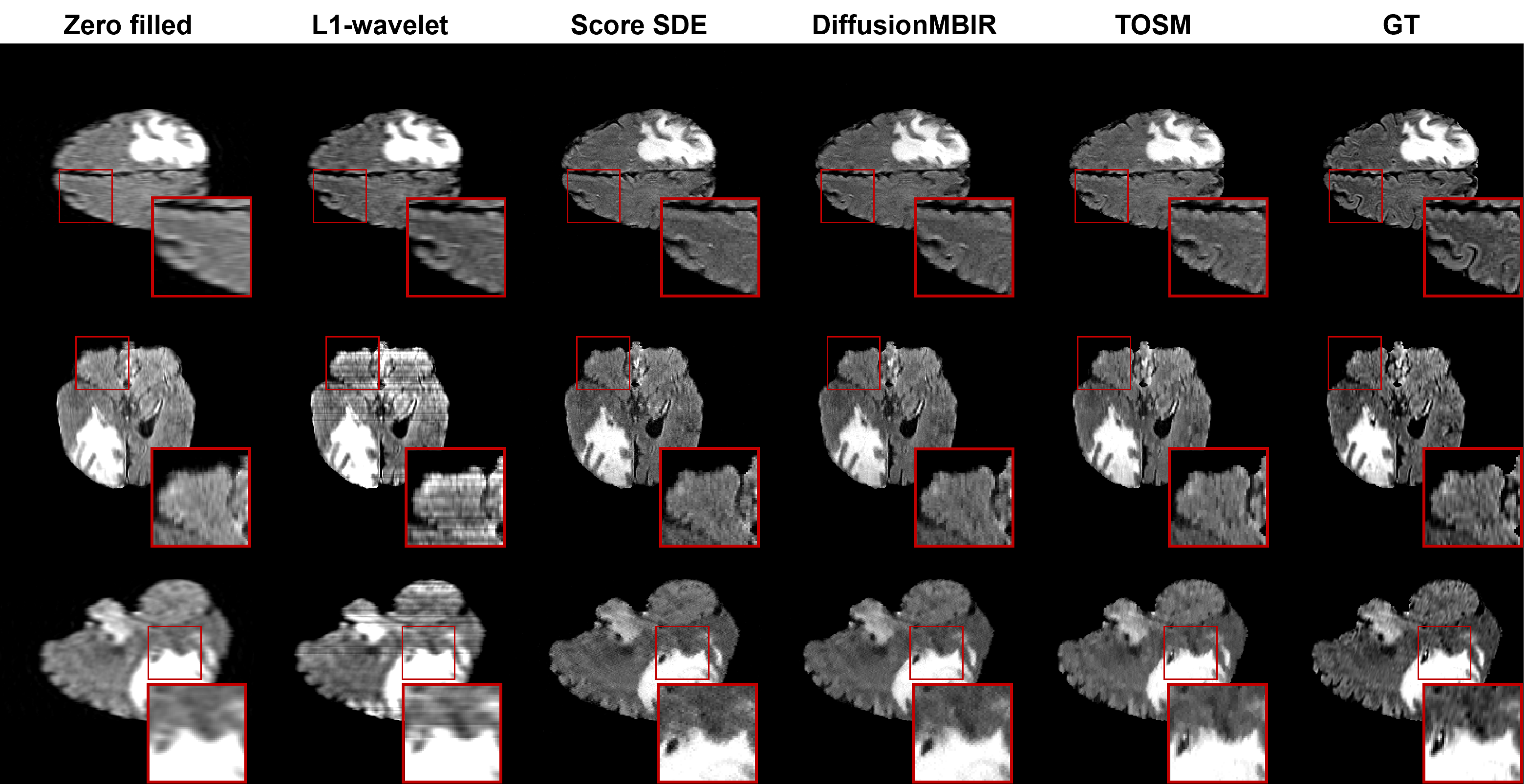}
    \caption{Representative results of fast MRI reconstruction from ×15 acceleration Gaussian 1D sub-sampled data. The first row is the primary plane, and the second row and third rows are auxiliary planes. The first to fifth columns correspond to different reconstruction methods, and the last column shows the GT images}
    \label{resultMRIx15}
\end{figure*}

\begin{figure}[t]
    \centering
    \includegraphics[scale=0.15]{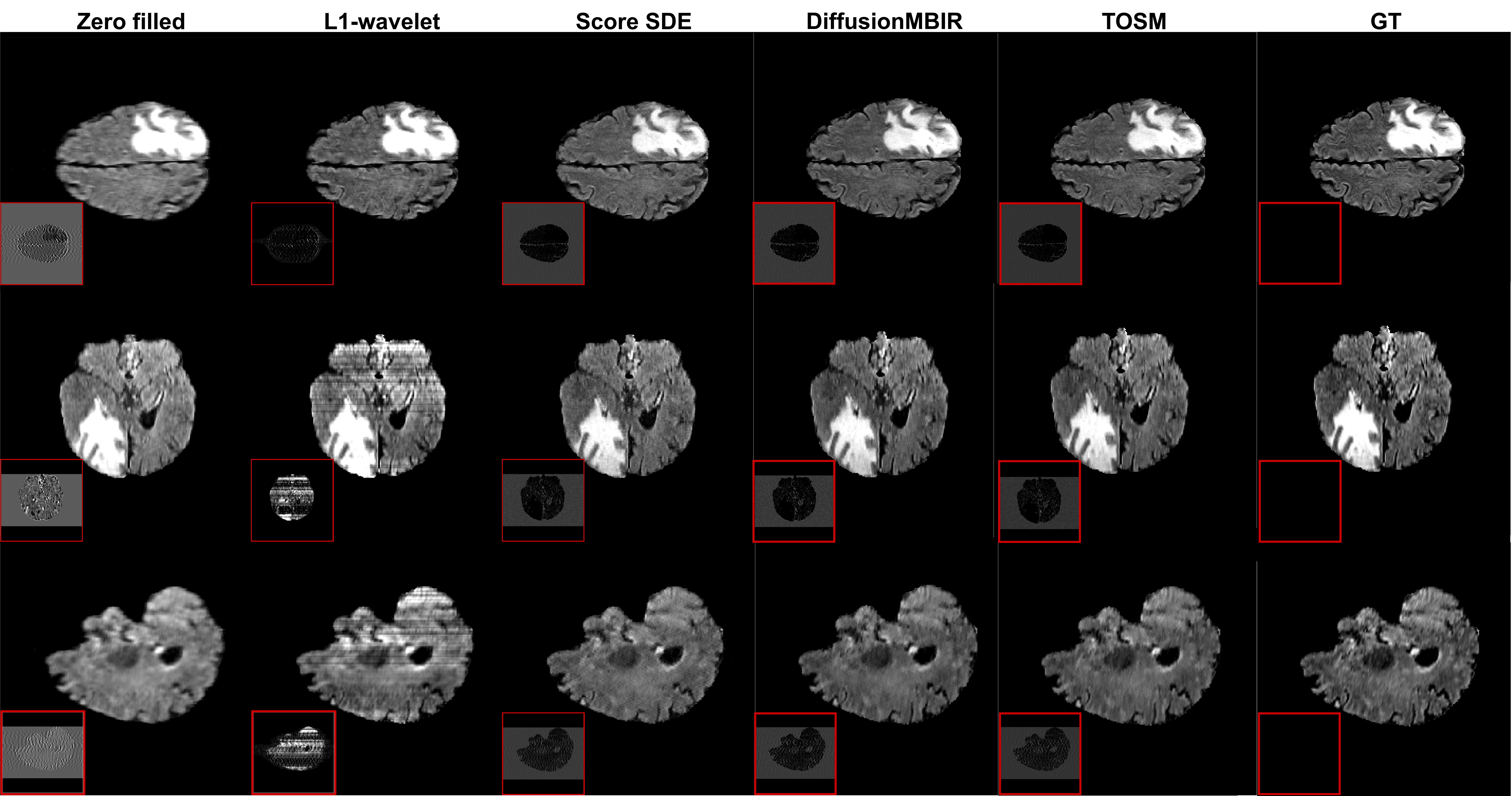}
    \caption{ Similar to Fig. \ref{resultMRIx15} but from ×2 acceleration Gaussian 1D sub-sampled data.}
    \label{resultMRIx2}
\end{figure}

\begin{figure}[ht]
    \centering
    \includegraphics[scale=0.3]{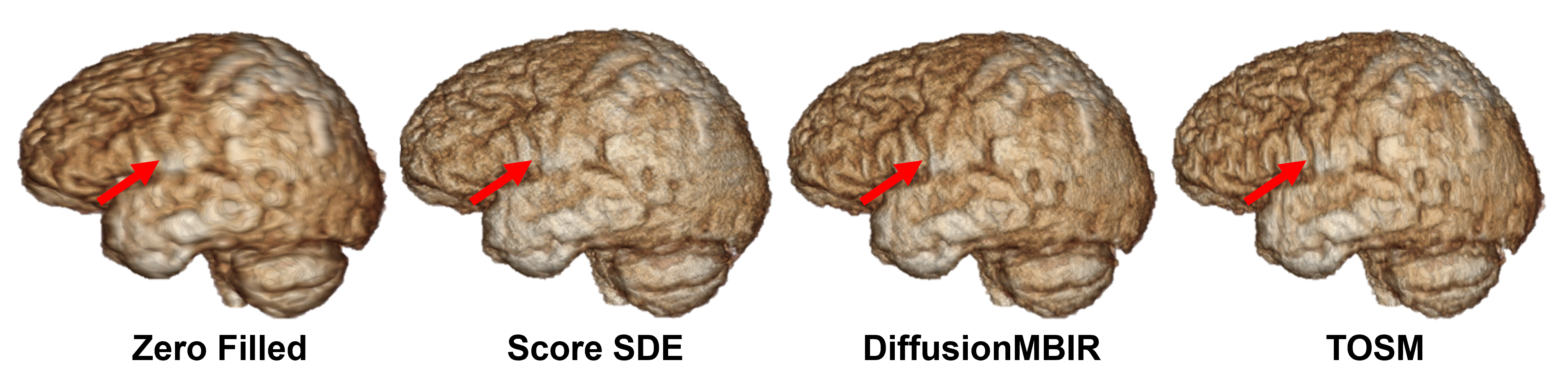}
    \caption{3D volumetric rendering of the brain MRI. The red arrows are used to highlight remarkable channel features within the brain.}
    \label{3dmri}
\end{figure}

\red{In this section, we present the experimental validation and demonstrate our investigation findings.} We begin by introducing the involved datasets. Then, we provide a comprehensive description of the key parameters and the experimental setup. Finally, we perform a comparative evaluation of our method against the state-of-the-art approaches, focusing on addressing the challenges of sparse-view CT and fast MRI reconstruction. Our method is open-sourced in GitHub \url{https://github.com/lizrzr/TOSM_3D-inverse-problem}.

\subsection{Datasets and Implementation Details}
\subsubsection{Datasets}
We utilize the Mayo Clinic abdomen images from the AAPM Low Dose CT Grand Challenge by McCollough \emph{et. al.} \cite{mccollough2016tu} for sparse-view CT reconstruction. The dataset \red{consists of} normal-dose CT scans from 10 patients, with 9 patients \red{used for} training and one patient for evaluation. The training dataset \red{contains} 4608 slices, \red{each with a thickness of 1.0 mm and a pixel grid of 512 × 512}. For the sparse-view testing set, we use cone-beam scanning geometry, where the distances from the rotation center to the source and detector are set as 50cm and 50cm, respectively. A flat panel detector is assumed \red{with a resolution of} $1024\times1024$ pixels, \red{where} each pixel covers an area of $0.08 \times 0.08$ mm². We generate only 29 views evenly distributed over 360 degrees for testing in this study.

In the MRI experiment, we utilize the publicly available fast MRI reconstruction dataset from the multi-modal brain tumor image segmentation benchmark (BRATS) 2018 \cite{menze2014multimodal}. We employ the pre-trained score function on fast MRI knee datasets \cite{zbontar2018fastmri}, \red{elimination} the need to process the MRI training data. Our training process follows the methodology outlined in \cite{songscore}. In the testing phase, we select brain images from a single patient with dimensions of $155\times240\times240$. By the zero-padding, we obtain \red{brain images of size} $240\times240\times240$ \red{pixels}.

\subsubsection{Model Training and Parameter Selection}
\red{To perform model training}, we \red{ultilize} the Adam algorithm with a learning rate of $10^{-3}$. Our method is implemented in Python, utilizing the ASTRA toolbox \cite{van2015astra} and PyTorch. The traditional score-based models for 3D volumetric reconstruction often require a high computational cost. \red{However, our method is capable of handling reconstruction tasks of size $512\times512\times512$ on a single RTX 3090 GPU.} \red{Therefore}, all experiments in this study are conducted using an RTX 3090 for computation.

To introduce a diverse range of noise levels, we generate twelve different noise levels. During the training phase, random noise corresponding to these twelve levels is added to each image and then fed into the network. \red{The training process iterates 100,000 times to ensure robust learning.} 
\red{The optimizer used is the Adam optimizer.} During the reconstruction phase, \red{we generate volumetric images at each of the twelve noise levels, employing 150 iterations for each level.} In each iteration, we enforce data consistency using the SIRT algorithm \red{with 20 iterations.} \red{To strike a balance between the generative capability of the score model and the constraint of data consistency, we assign weights of 0.5 to both factors.}

To quantitatively evaluate algorithm performance, \red{three standard evaluation metrics are computed: peak signal-to-noise ratio (PSNR), structural similarity (SSIM) and root mean squared error (RMSE).} We also generate visual representations of the reconstructed 3D volumetric images to compare their spatial and structural characteristics. Our goal is to demonstrate the effectiveness and superiority of our approach in addressing challenges related to sparse-view CT and fast MRI reconstruction. We rigorously follow this experimental setup to comprehensively evaluate and compare our method against existing techniques. 

\subsection{Sparse-view CT Reconstruction}
In this experiment, \red{various} comparison methods \red{were utilized, including} the traditional Feldkamp-Davis-Kress (FDK) \cite{van2015astra}, CNN-based FBPConvNet \cite{jin2017deep}, denoising score-based model (DSM) \cite{song2019generative} and DiffusionMBIR\cite{lee2023improving}. Particularly, the DiffusionMBIR is constrained with the DSM and total variation (TV). Figure \ref{resultCT} shows the reconstructed results from 29 projection views with different methods. 

The FDK algorithm, \red{as a classical analytic reconstruction method, generates transaxial, sagittal, and coronal images of inferior quality, with prominent streaky sparse-view artifacts}. Meanwhile, the FBPConvNet, trained solely on transaxial slices, only \red{manages to recover} rough image contours, resulting in the loss of fine details. The stacked sagittal and coronal images suffer from even more severe information loss with slice inconsistency. The DSM is one of the classical score-based reconstruction methods, by combining prior data distribution regularization and data consistency, it shows promising results in the transaxial plane but still has distortions in the details. In the stacked sagittal and coronal slices, significant inaccuracies \red{cause} distortions and errors in the magnified ROIs.
The DiffusionMBIR approach is designed to address the inter-slice \red{inconsistency in 3D reconstruction by employing a score-based model.} It incorporates TV in the z-direction to constrain the generation of sagittal and coronal planes. However, this approach does not explicitly model the 3D volumetric images, and the generated results are not as effective as our TOSM method. In contrast, our TOSM demonstrates \red{exceptional} results in 3D volumetric reconstruction. In the transaxial slice, the incorporation of complementary information from the sagittal and coronal views significantly improves image quality. \red{The reconstructed images minimize sparse-view artifacts and accurately restore details.} In the sagittal and coronal planes, our method stands out as the only one capable of producing accurate reconstructions.  The details in the ROIs precisely match the ground truths, demonstrating \red{exceptional performance in solving the inter-slice inconsistency issue} and highlighting the effectiveness of our approach.

Additionally, quantitative analysis is performed and the results are in Fig. \ref{violin} and Table \ref{tab1}. \red{We used three metrics: PSNR, SSIM, and RMSE. Higher PSNR and SSIM values indicate better reconstruction quality, while a smaller RMSE signifies a smaller gap between the reconstructed result and the ground truth. Remarkably, TOSM consistently achieved a 1.56 dB PSNR improvement over the existing state-of-the-art (SOTA) sparse-view CT 3D reconstruction method DiffusionMBIR. Besides, In both SSIM and RMSE metrics, TOSM also achieved the best results, indicating that the reconstructed 3D images using the TOSM method exhibit the closest structural similarity to the original 3D volume with minimal distortion.} Particularly, our TOSM achieves significant improvements in the sagittal and coronal planes compared to the transaxial plane. This observation verifies the superior 3D reconstruction capabilities of our approach over other methods.

Moreover, Figs. \ref{converge} and \ref{image_iter} demonstrate that the TOSM has fast convergence compared to the DSM and DiffusionMBIR. Obviously, the purple line representing TOSM is consistently above those of the DSM and DiffusionMBIR. With the same number of iterations, TOSM has a better performance. Besides, Fig. \ref{NPS} compares the noise power spectrum (NPS)\cite{kijewski1987noise}. It is seen that the DSM, DiffusionMBIR, and our TOSM significantly improve NPS compared to the other methods, evidenced by the star-shaped pattern observed in the plots. Especially, our TOSM yields the best superior results, outperforming the others by a considerable margin.

\red{Furthermore, to visually demonstrate the effectiveness of our method in 3D reconstruction tasks,} we present representative visualization results in Fig. \ref{fig3d}. It shows the connectivity of pulmonary vessels and smooth continuity of 3D structures, with noteworthy features indicated by red arrows. \red{It is evident that our TOSM results exhibit significantly stronger vascular connectivity compared to other methods, highlighting its exceptional ability to capture intricate details of pulmonary blood vessels. These results unequivocally establish the TOSM as the top approach for 3D volumetric reconstruction among the compared methods.}
\begin{table}[!h]
\centering
\caption{Quantitative Evaluation Results for Sparse-view CT Reconstruction}
\label{tab1}
\scalebox{0.65}{
\begin{tabular}{
>{\columncolor[HTML]{FFFFFF}}c 
>{\columncolor[HTML]{FFFFFF}}c 
>{\columncolor[HTML]{FFFFFF}}l 
>{\columncolor[HTML]{FFFFFF}}c 
>{\columncolor[HTML]{FFFFFF}}l 
>{\columncolor[HTML]{FFFFFF}}c 
>{\columncolor[HTML]{FFFFFF}}l }
\hline
{\color[HTML]{000000} \textbf{}} &
  \multicolumn{6}{c}{\cellcolor[HTML]{FFFFFF}\textbf{PSNR(dB) / SSIM / RMSE}} \\ \cline{2-7} 
\textbf{Method} &
  \multicolumn{2}{c}{\cellcolor[HTML]{FFFFFF}Transaxial*} &
  \multicolumn{2}{c}{\cellcolor[HTML]{FFFFFF}Sagittal} &
  \multicolumn{2}{c}{\cellcolor[HTML]{FFFFFF}Coronal+} \\ \hline
FDK &
  \multicolumn{2}{c}{\cellcolor[HTML]{FFFFFF}21.33 / 0.367 / 0.085} &
  \multicolumn{2}{c}{\cellcolor[HTML]{FFFFFF}29.71 / 0.264 / 0.032} &
  \multicolumn{2}{c}{\cellcolor[HTML]{FFFFFF}16.91 / 0.269 / 0.026} \\
FBPConvNet &
  \multicolumn{2}{c}{\cellcolor[HTML]{FFFFFF}32.76 / 0.872 / 0.023} &
  \multicolumn{2}{c}{\cellcolor[HTML]{FFFFFF}31.55 / 0.844 / 0.027} &
  \multicolumn{2}{c}{\cellcolor[HTML]{FFFFFF}29.08 / 0.844 / 0.035} \\
DSM &
  \multicolumn{2}{c}{\cellcolor[HTML]{FFFFFF}38.14 / 0.938 / 0.012} &
  \multicolumn{2}{c}{\cellcolor[HTML]{FFFFFF}36.82 / 0.917 / 0.014} &
  \multicolumn{2}{c}{\cellcolor[HTML]{FFFFFF}33.96 / 0.914 / 0.020} \\
DiffusionMBIR &
  \multicolumn{2}{c}{\cellcolor[HTML]{FFFFFF}38.52 / 0.943 / 0.012} &
  \multicolumn{2}{c}{\cellcolor[HTML]{FFFFFF}36.89 / 0.926 / 0.014} &
  \multicolumn{2}{c}{\cellcolor[HTML]{FFFFFF}34.56 / 0.916 / 0.019} \\
\textbf{TOSM} &
  \multicolumn{2}{c}{\cellcolor[HTML]{FFFFFF}{\color[HTML]{FF0000} \textbf{40.06}} / {\color[HTML]{FF0000} \textbf{0.954}} /
  {\color[HTML]{FF0000} \textbf{0.009}}} &
  \multicolumn{2}{c}{\cellcolor[HTML]{FFFFFF}{\color[HTML]{FF0000} \textbf{38.52}} / {\color[HTML]{FF0000} \textbf{0.933}} /
  {\color[HTML]{FF0000} \textbf{0.012}}} &
  \multicolumn{2}{c}{\cellcolor[HTML]{FFFFFF}{\color[HTML]{FF0000} \textbf{36.06}} / {\color[HTML]{FF0000} \textbf{0.922}} /
  {\color[HTML]{FF0000} \textbf{0.015}}} \\ \hline
\end{tabular}}
\end{table}
%%%%
\begin{table}[!h]
\centering
\caption{Quantitative Evaluation Results for Fast MRI Reconstruction}
\scalebox{0.65}{
\begin{tabular}{
>{\columncolor[HTML]{FFFFFF}}c 
>{\columncolor[HTML]{FFFFFF}}c 
>{\columncolor[HTML]{FFFFFF}}l 
>{\columncolor[HTML]{FFFFFF}}c 
>{\columncolor[HTML]{FFFFFF}}l 
>{\columncolor[HTML]{FFFFFF}}c 
>{\columncolor[HTML]{FFFFFF}}l }
\hline
{\color[HTML]{000000} \textbf{}} &
  \multicolumn{6}{c}{\cellcolor[HTML]{FFFFFF}\textbf{PSNR(dB) / SSIM / RMSE}} \\ \cline{2-7} 
\textbf{Method} &
  \multicolumn{2}{c}{\cellcolor[HTML]{FFFFFF}Transaxial*} &
  \multicolumn{2}{c}{\cellcolor[HTML]{FFFFFF}Sagittal} &
  \multicolumn{2}{c}{\cellcolor[HTML]{FFFFFF}Coronal+} \\ \hline
    \multicolumn{7}{c}{\cellcolor[HTML]{FFFFFF} FASTMRI X15 GAUSSIAN 1D
RECONSTRUCTION} \\ \hline
Zero-filled & 
  \multicolumn{2}{c}{\cellcolor[HTML]{FFFFFF}27.79 / 0.712 / 0.041} &
  \multicolumn{2}{c}{\cellcolor[HTML]{FFFFFF}25.87 / 0.551 / 0.051} &
  \multicolumn{2}{c}{\cellcolor[HTML]{FFFFFF}27.78 / 0.711 / 0.041} \\
L1-wavelet &
  \multicolumn{2}{c}{\cellcolor[HTML]{FFFFFF}29.88 / 0.664 / 0.032} &
  \multicolumn{2}{c}{\cellcolor[HTML]{FFFFFF}27.85 / 0.617 / 0.040} &
  \multicolumn{2}{c}{\cellcolor[HTML]{FFFFFF}29.73 / 0.661 / 0.033} \\
Score SDE &
  \multicolumn{2}{c}{\cellcolor[HTML]{FFFFFF}31.54 / 0.815 / 0.027} &
  \multicolumn{2}{c}{\cellcolor[HTML]{FFFFFF}29.54 / 0.708 / 0.033} &
  \multicolumn{2}{c}{\cellcolor[HTML]{FFFFFF}31.41 / 0.812 / 0.027} \\
DiffusionMBIR &
  \multicolumn{2}{c}{\cellcolor[HTML]{FFFFFF}32.27 / 0.829 / 0.025} &
  \multicolumn{2}{c}{\cellcolor[HTML]{FFFFFF}30.11 / 0.730 / 0.032} &
  \multicolumn{2}{c}{\cellcolor[HTML]{FFFFFF}32.01 / 0.827 / 0.025} \\
\textbf{Our} &
  \multicolumn{2}{c}{\cellcolor[HTML]{FFFFFF}{\color[HTML]{FF0000} \textbf{32.72}} / {\color[HTML]{FF0000} \textbf{0.846}} / 
  {\color[HTML]{FF0000} \textbf{0.023}}} &
  \multicolumn{2}{c}{\cellcolor[HTML]{FFFFFF}{\color[HTML]{FF0000} \textbf{30.67}} / {\color[HTML]{FF0000} \textbf{0.760}} /
  {\color[HTML]{FF0000} \textbf{0.029}}} &
  \multicolumn{2}{c}{\cellcolor[HTML]{FFFFFF}{\color[HTML]{FF0000} \textbf{32.55}} / {\color[HTML]{FF0000} \textbf{0.844}} /
  {\color[HTML]{FF0000} \textbf{0.024}}} \\ \hline
    \multicolumn{7}{c}{\cellcolor[HTML]{FFFFFF} FASTMRI X2 UNIFORM 1D
RECONSTRUCTION} \\ \hline
Zero-filled &
  \multicolumn{2}{c}{\cellcolor[HTML]{FFFFFF}29.95 / 0.787 / 0.032} &
  \multicolumn{2}{c}{\cellcolor[HTML]{FFFFFF}27.98 / 0.679 / 0.040} &
  \multicolumn{2}{c}{\cellcolor[HTML]{FFFFFF}29.90 / 0.788 / 0.032} \\
L1-wavelet &
  \multicolumn{2}{c}{\cellcolor[HTML]{FFFFFF}30.57 / 0.757 / 0.032} &
  \multicolumn{2}{c}{\cellcolor[HTML]{FFFFFF}29.31 / 0.707 / 0.034} &
  \multicolumn{2}{c}{\cellcolor[HTML]{FFFFFF}30.49 / 0.751 / 0.030} \\
Score SDE &
  \multicolumn{2}{c}{\cellcolor[HTML]{FFFFFF}34.78 / 0.893 / 0.018} &
  \multicolumn{2}{c}{\cellcolor[HTML]{FFFFFF}32.86 / 0.830 / 0.023} &
  \multicolumn{2}{c}{\cellcolor[HTML]{FFFFFF}34.76 / 0.892 / 0.018} \\
  DiffusionMBIR &
  \multicolumn{2}{c}{\cellcolor[HTML]{FFFFFF}34.99 / 0.901 / 0.017} &
  \multicolumn{2}{c}{\cellcolor[HTML]{FFFFFF}33.12 / 0.838 / 0.022} &
  \multicolumn{2}{c}{\cellcolor[HTML]{FFFFFF}34.98 / 0.896 / 0.017} \\
\textbf{Our} &
  \multicolumn{2}{c}{\cellcolor[HTML]{FFFFFF}{\color[HTML]{FF0000} \textbf{35.95}} / {\color[HTML]{FF0000} \textbf{0.915}} /
  {\color[HTML]{FF0000} \textbf{0.015}}} &
  \multicolumn{2}{c}{\cellcolor[HTML]{FFFFFF}{\color[HTML]{FF0000} \textbf{33.93}} / {\color[HTML]{FF0000} \textbf{0.864}} /
  {\color[HTML]{FF0000} \textbf{0.020}}} &
  \multicolumn{2}{c}{\cellcolor[HTML]{FFFFFF}{\color[HTML]{FF0000} \textbf{35.83}} / {\color[HTML]{FF0000} \textbf{0.913}} /
  {\color[HTML]{FF0000} \textbf{0.016}}} \\ \hline
\end{tabular}}
\label{mrilabelx15}
\end{table}
\subsection{Fast MRI Reconstruction}
\red{To comprehensively evaluate the effectiveness of our method in undersampled MRI reconstruction, we conducted experiments on 1D uniformly and Gaussian 1D random sub-sampling scenarios.} Specifically, we retain only 15\% of the auto-calibration signal (ACS) region at the center and sample only half of the k-space lines, resulting in acceleration factors of approximately ×2 and ×15, respectively.

The results of the ×15 accelerated reconstruction, as depicted in Fig. \ref{resultMRIx15}, demonstrate remarkable similarities to high-quality CT scans. Particularly, the transaxial plane showcases the most favorable reconstruction performance. In comparison to the L1-wavelet algorithm \cite{doi:10.1073/pnas.2201062119}, which introduces severe discontinuities and artifacts in the other two planes, our approach exhibits superior performance compared to 2D score-based models. Notably, our method achieves more accurate details in the coronal and sagittal planes when compared to the current baseline method DiffusionMBIR. The quantitative statistical results in Table \ref{mrilabelx15} further confirm the superior reconstruction quality of our proposed TOSM method in comparison to other model-based methods. \red{Specifically, we evaluated the PSNR, SSIM, and RMSE metrics in the sagittal, coronal, and transaxial planes. The table results indicate that our method outperforms the current state-of-the-art 3D reconstruction method, DiffusionMBIR, across all three metrics.}

Additionally, the results of ×2 acceleration (see Fig. \ref{resultMRIx2}) exhibit high-fidelity reconstructions on all three anatomical planes. Conversely, the L1-wavelet method encounters severe discontinuities and artifacts in the other two planes. Again, the quantitative statistical results in Table \ref{mrilabelx15} support the conclusion that our TOSM method outperforms other comparison methods, \red{averaging a 0.87 dB
PSNR improvement over existing 3D fast MRI reconstruction method DiffusionMBIR.} Furthermore, we provide a 3D rendering of the MRI reconstruction result in Fig. \ref{3dmri}, focusing on the region indicated by red arrows. Notably, the contours generated by our TOSM method exhibit greater clarity and accuracy, while the other methods have different degrees of distortions.

The experimental results offer compelling evidence for the effectiveness and generalizability of our approach in fast MRI reconstruction tasks. By effectively addressing the reconstruction challenges associated with downsampled k-space data, our TOSM method outperforms existing algorithms, resulting in high-quality reconstructions that faithfully preserve essential structural information in 3D volumetric images.
\red{\section{Discussion}}
\red{
\textbf{Outlook of Expanding to higher-order dimensions:} In the TOSM method, we employ a global score update strategy on the 3D volume using a pre-trained 2D score-based model, which was crucial for achieving 3D image reconstruction. We demonstrated the feasibility of 3D imaging for medical images using our TOSM. Theoretically, 
when the target data distribution is similar across different dimensions, our method can be extended to higher-order. Certainly, when extending TOSM to higher-order, there may be some errors introduced. In our future work, we will continue to explore and address this challenge.}

\red{\textbf{Limitation of Approximated Distribution:} When constructing the theoretical approach of TOSM, we observed that different directional planes of the 3D volume exhibit similar distributions in structural features and grayscale histograms. Therefore, a single pre-trained score-based model was used to calculate scores in these three directions. While this observation aligns with intuition, our forthcoming work will focus on exploring whether this approximation is mathematically bounded.}
\section{Conclusion}
In this study, the 2.5-order score-based model (TOSM) marks a significant stride in addressing 3D ill-posed inverse problems. \red{TOSM showcases an exceptional ability to precisely compute 3D scores by utilizing a single pre-trained 2D score-based model.} This distinctive approach enables streamlined global updates of 3D volumetric images while circumventing substantial computational overhead.

\red{This research unfolds substantial practical implications.} The experimental findings consistently validate the efficacy and adaptability of our method across diverse tasks, spanning sparse-view CT reconstruction on the AAPM datasets to fast MRI reconstruction with ×2 and ×15 acceleration. \red{Whether in terms of quantitative metrics like PSNR, SSIM, or in image quality, TOSM achieved state-of-the-art results.} Notably, our approach excels in tackling the challenge of inter-slice inconsistency, manifesting exceptional performance in resolving typical 3D inverse problems.

\red{More exploration is needed in the future work. In the process of using TOSM to address the inverse problem of 3D reconstruction, we inferred the feasibility of estimating the distribution in three directions using a single score-based model by observing the similarity between the grayscale histogram distribution and the detailed structure in three directions. However, this approximation might require mathematical proof in future work.}

\red{In future work, we plan to undertake two further studies. First, employing mathematical tools to demonstrate that the error of estimating the data distribution in three directions using a single score-based model is acceptable. Second, to expand the TOSM method to higher dimensions, such as utilizing pre-trained 3D score-based model to handle 4D signals like temporal CT data.}
\bibliographystyle{unsrt}
% Loading bibliography database
\bibliography{cas-refs}

% Biography
% \bio{}
% % Here goes the biography details.
% \endbio

% \bio{pic1}
% % Here goes the biography details.
% \endbio

\end{document}